\documentclass[useamsfonts]{pasj01}

\usepackage{graphicx,todonotes}
\usepackage{color,xcolor}
\usepackage{amssymb}
\usepackage[fleqn]{amsmath}
\usepackage{natbib}

\newcommand{\simgt}{\lower.5ex\hbox{$\; \buildrel > \over \sim \;$}}
\newcommand{\simlt}{\lower.5ex\hbox{$\; \buildrel < \over \sim \;$}}

\def\mhunit{\mbox{$10^{14}\,M_\odot/h$}}

\def\mvir{\mbox{$M_\mathrm{200c}$}}
\def\cvir{\mbox{$c_\mathrm{200c}$}}

\def\mpch{\mbox{$\mathrm{Mpc}/h$}}
\def\kpch{\mbox{$\mathrm{kpc}/h$}}
\def\lcdm{\mbox{$\Lambda$CDM}}

\def\sig8{\mbox{$\sigma_8$}}
\def\Sigmacr{\mbox{$\Sigma_{\rm cr}$}}
\def\DSigma{\mbox{$\Delta\Sigma$}}
\def\ncor{\mbox{$N_{\rm mem}$}}
\def\equationautorefname~#1\null{Equation~(#1)\null}

\begin{document}
\title{Source Selection for Cluster Weak Lensing Measurements in the Hyper Suprime-Cam Survey} 

\author{Elinor  {Medezinski}\altaffilmark{1}}  
\author{Masamune {Oguri}\altaffilmark{2,3,4}}
\author{Atsushi~J. {Nishizawa}\altaffilmark{5}}
\author{Joshua~S. {Speagle}\altaffilmark{6}}
\author{Hironao {Miyatake}\altaffilmark{2,7}}
\author{Keiichi {Umetsu}\altaffilmark{8}}
\author{Alexie {Leauthaud}\altaffilmark{9}}
\author{Ryoma {Murata}\altaffilmark{2,4}}
\author{Rachel {Mandelbaum}\altaffilmark{10}}
\author{Crist\'obal {Sif\'on}\altaffilmark{1}}
\author{Michael~A. {Strauss}\altaffilmark{1}}
\author{Song {Huang}\altaffilmark{2,9}}
\author{Melanie {Simet}\altaffilmark{7,11}}
\author{Nobuhiro {Okabe}\altaffilmark{12,13}}
\author{Masayuki {Tanaka}\altaffilmark{14}}
\author{Yutaka {Komiyama}\altaffilmark{14,15}}

\email{elinorm@astro.princeton.edu}

\altaffiltext{1}{Department of Astrophysical Sciences, Princeton University, 4 Ivy Lane, Princeton, NJ 08544, USA} 
\altaffiltext{2}{Kavli Institute for the Physics and Mathematics of the Universe (Kavli IPMU, WPI), TokyoInstitutes for Advanced Study, The University of Tokyo, Chiba 277-8582, Japan}
\altaffiltext{3}{Research Center for the Early Universe, University of Tokyo, Tokyo 113-0033, Japan}
\altaffiltext{4}{Department of Physics, University of Tokyo, Tokyo 113-0033, Japan}
\altaffiltext{5}{Institute for Advanced Research, Nagoya University, Nagoya 464-8602, Aichi, Japan}
\altaffiltext{6}{Harvard University, 60 Garden St, Cambridge, MA 02138}
\altaffiltext{7}{Jet Propulsion Laboratory, California Institute of Technology, Pasadena, CA 91109, USA}
\altaffiltext{8}{Institute of Astronomy and Astrophysics, Academia
Sinica, P.~O. Box 23-141, Taipei 10617, Taiwan}  
\altaffiltext{9}{Department of Astronomy and Astrophysics, University of California, Santa Cruz, 1156 HighStreet, Santa Cruz, CA 95064 USA}
\altaffiltext{10}{McWilliams Center for Cosmology, Department of Physics, Carnegie Mellon University,Pittsburgh, PA 15213, USA}
\altaffiltext{11}{University of California, Riverside, 900 University Avenue, Riverside, CA 92521, USA}
\altaffiltext{12}{Department of Physical Science, Hiroshima University, 1-3-1 Kagamiyama,Higashi-Hiroshima, Hiroshima 739-8526, Japan}
\altaffiltext{13}{Hiroshima Astrophysical Science Center, Hiroshima University, Higashi-Hiroshima,Kagamiyama 1-3-1, 739-8526, Japan}
\altaffiltext{14}{National Astronomical Observatory of Japan, 2-21-1 Osawa, Mitaka, Tokyo 181-8588, Japan}
\altaffiltext{15}{Department of Astronomy, School of Science, SOKENDAI (The Graduate University for Advanced Studies), 2-21-1 Osawa, Mitaka,  Tokyo 181-8588, Japan}

\maketitle


\begin{abstract}

We present  optimized source galaxy selection schemes for measuring   cluster weak lensing (WL) mass profiles unaffected by cluster member dilution from the Subaru Hyper Suprime-Cam Strategic Survey Program (HSC-SSP). 
The ongoing HSC-SSP survey will uncover thousands of galaxy clusters to $z\lesssim1.5$. In deriving cluster masses via WL, a critical source of systematics is contamination and dilution of the lensing signal by cluster {members, and by foreground galaxies whose photometric redshifts are biased}. Using the first-year CAMIRA catalog of $\sim$900 clusters with richness larger than 20 found  in $\sim$140~deg$^2$ of HSC-SSP data,  we devise and compare several source selection methods, including selection in color-color space (CC-cut), and selection of robust photometric redshifts by applying constraints on their cumulative  probability distribution function (PDF; P-cut). We examine the dependence of the contamination on the chosen limits adopted for each method. Using the proper limits, these methods give  mass profiles with minimal dilution in agreement with one another. We find that not adopting either the CC-cut or P-cut methods results in  an underestimation of the total cluster mass ($13\pm4 \%$) and the concentration of the profile ($24\pm11\%$). 
The level of cluster contamination can reach as high as $\sim10$\% at $R\approx 0.24$~\mpch~ for low-z clusters without cuts, while employing either the P-cut or CC-cut  results in cluster contamination  consistent with zero to  within the 0.5\% uncertainties. 
Our robust methods yield a $\sim60\sigma$ detection of the stacked CAMIRA  surface mass density profile, with a mean mass of $\mvir = (1.67\pm0.05({\rm {stat}}))\times\mhunit$.
\end{abstract}
 

\section{Introduction} 
\label{sec:intro}
Tracing the exponential tail of the halo mass function, clusters of galaxies are a powerful probe of cosmology, and in particular,  their abundance is sensitive to the late-time nonlinear growth of structure. Placing competitive cosmological constraints with cluster abundances requires precise and accurate masses for these objects. Calibrations of indirect mass proxies for clusters detected by the Sunyaev-Zel'dovich (SZ) effect \citep{Sunyaev1972}, X-ray or  optical surveys typically rely on scaling relations calibrated via alternative methods. Some of these relations make  assumptions about the cluster dynamical state, e.g.,  hydrostatic equilibrium (HSE) in the case of  X-ray observations.

The distribution of mass within clusters provides further insight into dark matter (DM) and structure formation scenarios.  Simulations of cold DM (CDM) dominated halos consistently predict mass profiles that steepen with radius, providing a distinctive, fundamental prediction for this form of DM \citep{NFW97}. Furthermore, the degree of mass concentration, $c_{\rm vir} = r_{\rm vir}/r_{\rm s}$, the ratio of the virial radius $r_{\rm vir}$ to the inner characteristic radius $r_{\rm s}$, should decline with increasing cluster mass as the more massive clusters collapse later when the cosmological background density is lower. A precise determination of the inner ($<$200 kpc) density slope of DM halos is of great importance for DM annihilation experiments \citep{Su2012} .

The best direct probe of the total mass and its (projected) distribution in clusters is via weak gravitational lensing (WL), as it requires no assumption for the dynamical state of the cluster or the nature of DM. 
WL gives rise to the coherent distortion of galaxy shapes, measured statistically over thousands of background galaxies. {Since lensing is only sensitive to the projected matter density, the triaxiality of cluster halos leads to an intrinsic scatter of 15--20\%  for the mass of each cluster when compared to other methods \citep{Oguri2005,Corless2007,Meneghetti2010,Becker2011}.
With current samples of hundreds of clusters, cluster mass profiles can be stacked to enhance the signal-to-noise ratio (S/N) and reduce the scatter due to triaxiality \citep{Umetsu2014,Okabe2016,Simet2017}. }This way,  cluster mass calibration has reached   $\sim6$--$10\%$  precision at intermediate redshifts and masses \citep{von-der-Linden2014,Hoekstra2015,Penna-Lima2017,Smith2016,Melchior2017,Simet2017}.

Several systematic effects inherent to WL can {bias the cluster mass calibration}. 
Instrumental and observational distortions can cause systematic signals that are similar in size to the gravitational distortion of galaxy shapes \citep{Mandelbaum2005a}. Following recent extensive tests, current techniques can now reach an accuracy of 1--2\% in shape measurement, following proper calibration with image simulations \citep{Heymans2006,Massey2007,Bridle2010,Kitching2012,Mandelbaum2015}. 

However, a  major source of systematics comes in correctly estimating the redshift distribution of source galaxies lying behind the clusters,  required to convert the lensing signal into a physical mass. Contamination by unlensed cluster and foreground galaxies causes a systematic underestimation of the true lensing mass profile \citep{Broadhurst2005a}. In particular, inclusion of cluster galaxies significantly  dilutes the signal closer to the cluster center  and causes an underestimation of the concentration of the density profile. {In contrast, the inclusion of foreground galaxies in the background source sample due to photometric redshift errors produces a dilution of the cluster lensing signal that does not depend on the cluster-centric radius. In this paper, we investigate both types of contamination of the source sample.}

Acquiring spectroscopic redshifts (spec-z's) for each source is not feasible, particularly to the depths WL observations now reach. Photometric redshifts (photo-z's) are typically used instead, but until recently,  cluster lensing studies relied on at most two to three observed bands \citep[e.g.,][]{Medezinski2010,Okabe2010,Oguri2012}, so that  reliable photo-z's  could not be determined. In turn, well-calibrated  field photo-z catalogs such as COSMOS \citep{Ilbert2009} were used to determine the redshift distribution \citep{Medezinski2010,Umetsu2012,Okabe2010}. Such field surveys are often limited to deep, small areas, and are subject to significant cosmic variance. Furthermore, this approach does not correct for contamination of the source sample by cluster galaxies. The enhancement of lens-source pair counts relative to random (known as ``boost'' factor) are used to correct for cluster contamination \citep{Sheldon2004,Hoekstra2007}, but those can be unreliable if the cluster sample is small \citep{Simet2017a} or if there is not enough spatial coverage to make use of fields adjacent to the cluster \citep{Applegate2014}. Additionally, magnification bias \citep{Umetsu2014,Chiu2016,Ziparo2016}, masking by cluster galaxies \citep{Simet2015}, and galaxy  selection effects need to be carefully accounted for when determining the boost \citep{Mandelbaum2006b}.

In the coming decade, many wide optical surveys are aimed at constraining cosmology via WL, e.g., the ongoing   Dark Energy Survey \citep[DES;][]{The-Dark-Energy-Survey-Collaboration2005}, the Kilo Degree Survey \citep[KiDS;][]{de-Jong2013}, and the Hyper Suprime-Cam \citep[HSC;][]{Aihara2017,Aihara2017a} survey, and the planned  Large Synoptic Survey Telescope \citep[LSST;][]{LSST}. These will observe in four to six broad bands,  so that the photo-z's are  better inferred, to a mean level of $\lesssim5\%$. However, these photo-z's will still be plagued by a large fraction of outliers ($\sim20-30\%$) due to inherent color-redshift degeneracies.
These degeneracies stem from having a finite number of broad optical bands  that do not span a wide enough wavelength range, particularly ultraviolet or infrared \citep{Benitez2009,Rafelski2009}. Another complication in the case of template-fitting codes \citep[e.g.,][]{Benitez2000} is that the template libraries often may not include the full range of galaxy spectral energy distribution (SED) features, e.g., accounting for emission lines or dust obscuration. 
The photo-z uncertainties are folded in by incorporating the full probability distribution function (PDF) of the individual photo-z's \citep{Applegate2014}. However, the PDFs suffer from large dependency on the assumed priors, and the representability of the  reference spec-z sample used for training.
Other approaches rely on more stringent color cuts to reject outlier photo-z's \citep{Medezinski2010,Okabe2010}, but then suffer from lower statistical power, as they result in lower source densities.

In this paper, we aim to explore the  systematic effects cluster and foreground contaminations have on cluster WL studies by using the exquisite deep data from  the Hyper Suprime-Cam Strategic Survey Program (HSC-SSP; Miyazaki et al. 2017, Aihara et al. 2017a,b). HSC-SSP, an ongoing survey in five optical bands ($grizy$), will reach unprecedented depths ($i\sim26$) for a large area ($1400\deg^2$ when finished). About a thousand clusters have already been identified to $z\lesssim1.1$ in its currently observed $240\deg^2$ HSC-Wide fields \citep{Oguri2017} using the red-sequence-based cluster finding algorithm, CAMIRA \citep{Oguri2014}. {The stacked CAMIRA cluster lensing signal will provide a 4\% (7\%) statistical constraint on the mean cluster mass at low (high) redshifts.} Here we make use of this large statistical cluster sample and test several source selection methods that optimize the use of robust photo-z's and minimize the contamination by cluster galaxies {in order to reduce the systematic level below that required from statistics.}

This paper is organized as follows. In Section~\ref{sec:WLmethod} we present the basic WL methodology. In Section~\ref{sec:data} we present the HSC survey, the dataset and the CAMIRA cluster catalog derived from HSC. In Section~\ref{sec:SS} we present the source selection methods considered in this paper. In Section~\ref{sec:WLprof} we present the resulting  mass profiles derived using the different selection methods and their biases as inferred from modeling. In Section~\ref{sec:tests} we present validation tests on the level of contamination in each method, and we summarize and conclude in Section~\ref{sec:summary}. Throughout this paper we adopt a {\it WMAP9} \citep{Hinshaw2013} $\lcdm$ cosmology, where $\Omega_M=0.282,\ \Omega_\Lambda=0.718$, and $h=H_0/100$ km~s$^{-1}$ Mpc$^{-1}$.

\section{Weak Lensing Methodology}
\label{sec:WLmethod}
Weak lensing  distorts  source galaxy shapes.  The amplitude of this distortion is proportional to all matter contained in the lensing cluster and along the line of sight to the lens.
The tangential distortion profile is related to the projected surface-mass density profile of the average mass distribution around the cluster,
\begin{equation} \label{eq:shear}
\gamma_T(R) = \frac{\DSigma(R)}{\Sigmacr} =  \frac{\bar{\Sigma}(<R)-\Sigma(R)}{\Sigmacr},
\end{equation}
where $R$ is the comoving transverse separation between the source and the lens, $\Sigma(R)$ is the projected surface mass density, $\bar{\Sigma}(<R)$ is the mean density within $R$, and
\begin{equation}\label{eq:sigcrit}
\Sigmacr =  \frac{c^2}{4\pi G}\frac{D_A(z_s)}{D_A(z_l)D_A(z_l,z_s)(1+z_l)^2},
\end{equation}
is the critical surface mass density, where $G$ is the gravitational constant, $c$ is the speed of light, $z_l$ and $z_s$ are the lens and source redshifts, respectively, and $D_A(z_l)$, $D_A(z_s)$,   and $D_A(z_l,z_s)$ are the angular diameter distances to the lens, the source, and the lens-to-source, respectively. The extra factor of  $(1+z_l)^2$ comes from our use of comoving coordinates \citep{Bartelmann2001a}. 

We estimate the mean projected mass density excess profile $\DSigma(R)$ from Equation~\ref{eq:shear} by stacking the shear over a population of source galaxies $s$ over multiple clusters $l$ that lie within a given cluster-centric radial annulus $R$,
\begin{equation}\label{eq:DSigma}
\DSigma(R) = \frac{1}{2\mathcal{R}(R)}\frac{ \sum\limits_{l,s} w_{ls} e_{t,ls}[\langle\Sigmacr^{-1}\rangle_{ls}]^{-1} } {(1+K(R)) \sum\limits_{l,s}w_{ls}},
\end{equation}
where the double summation is over all clusters and over all sources associated with each cluster (i.e., lens-source pairs). In the above expression, the measured  tangential shape distortion of a source galaxy is
\begin{equation}\label{eq:et}
e_t= -e_1 \cos{2\phi} - e_2 \sin{2\phi},
\end{equation}
where $\phi$ is the angle measured in sky coordinates from the right ascension direction to a line connecting the lens and source galaxy, and $e_1,e_2$ are the shear components in sky coordinates  obtained from the pipeline (see below). The $45\degree$-rotated component, $e_\times$, is also similarly computed, and is expected to be zero. The mean critical density  $\langle\Sigma_{{\rm cr}}^{-1}\rangle_{ls}^{-1}$   is averaged with the source photo-z PDF, $P(z)$, for each lens-source pair, such that
\begin{equation}\label{eq:sigcritmean}
\langle\Sigmacr^{-1}\rangle_{ls} = \frac{ \int_{z_l}^{\infty} \Sigmacr^{-1}(z_l,z) P(z)\mathrm{d}z} { \int_{0}^{\infty} P(z)\mathrm{d}z}.
\end{equation}
As long as the mean $P(z)$ correctly describes the sample redshift distribution, the above 
equation  corrects for  dilution  by cluster or foreground source galaxies. However, with limited wide optical bands, this is not always achievable, as we show below.
The weight in Equation~\ref{eq:DSigma}, $w_{ls}$, is given by
\begin{equation}
  \label{eq:lensweight}
w_{ls} = (\langle\Sigmacr^{-1}\rangle_{ls})^2 \frac{1}{\sigma_{e,s}^2+e_{{\rm rms},s}^2},
\end{equation}
where $\sigma_e$ is the per-component shape measurement uncertainty, and $e_{\rm rms}\approx0.40$ is the  RMS ellipticity estimate per component. The factor $(\langle\Sigmacr^{-1}\rangle_{ls})^2$ downweights pairs that are close in redshift to the lens.
The factor $(1 + K(R))$ in Equation~\ref{eq:DSigma} corrects for a multiplicative shear bias $m$ as determined from the GREAT3-like simulations \citep{Mandelbaum2014,Mandelbaum2015} and is described in Mandelbaum et al (in prep.). It is calculated as
\begin{equation}
K(R) = \frac{\sum_{l,s}m_{s}w_{ls}}{\sum_{l,s}w_{ls}}.
\end{equation}
The  `shear responsivity' factor in Equation~\ref{eq:DSigma}, 
\begin{equation}
\mathcal{R}(R) = 1-\frac{\sum_{l,s} e^2_{\rm rms,s}w_{ls}}{\sum_{l,s} w_{ls}}\approx 0.84, 
\end{equation}
represents the response of the ellipticity, $e$, to a small shear \citep{Kaiser1995,Bernstein2002}. A full description and  clarification of this procedure is given in \cite{Mandelbaum2017}.
Finally, the covariance matrix includes the statistical uncertainty due to shape noise,
\begin{equation}\label{eq:cov}
C^{\rm stat}(R)=\frac{1}{4{\cal R}^2(R)}\frac{\sum_{l,s} {w}^2_{ls} (e^2_{{\rm rms},s} + \sigma_{e, s}^2)\left\langle \Sigma_{{\rm cr}}^{-1} \right\rangle_{ls}^{-2}}{\left[1+K(R)\right]^2\left[ \sum_{l,s} {w}_{ls}\right]^2}.
\end{equation}
{We only include in the covariance the statistical uncertainty due to shape noise. While other sources of uncertainty should be considered, e.g. due to uncorrelated large-scale structures \citep{Hoekstra2003}, photo-z's and  the shear multiplicative bias correction, in this paper we are only interested in comparing the systematic error due to source selection with the uncertainty induced by statistics, rather that present a full mass calibration of the CAMIRA clusters.}

\section{Data}
\label{sec:data}
\subsection{HSC Observations}

The HSC-SSP \citep{Aihara2017a} is conducting an optical imaging survey with the new  1.77 deg$^2$ HSC camera (Miyazaki et al. 2017) installed on the Subaru 8.2m Telescope. The survey is designed  to have three layers: Wide, Deep and UltraDeep. The Wide survey, when completed, will span  $\sim$1400 deg$^2$. For this study, we  use the first 140 deg$^2$ of full-depth full-color (FDFC) data of the S16A internal data release. It is incremental to the first public data release, S15B, presented in \cite{Aihara2017}. The Wide layer consists of five broad bands, $grizy$, reaching a typical limiting magnitude of $i\sim26$, and exceptional mean seeing of FWHM$=0.6\arcsec$ in the $i$ band. The HSC data are reduced with the HSC Pipeline, \texttt{hscPipe} \citep{Bosch2017}, which is based on the LSST pipeline \citep{LSST,Axelrod2010,Juric2015}.
Seven different photo-z codes have been implemented by the team  \citep{Tanaka2017}.

The WL catalogs  derived from the HSC observations are described in detail in \cite{Mandelbaum2017}. In short, galaxy shapes are measured from the coadded $i$-band images using the re-Gaussianization method \citep{Hirata2003} that was extensively used and tested in the Sloan Digital Sky Survey \citep[SDSS;][]{Mandelbaum2005,Reyes2012,Mandelbaum2013} and its performance for HSC has been further characterized in \cite{Mandelbaum2017}. {The description of the shape catalog, its properties and the cuts applied are further described in Section~\ref{subsec:SSall}.}

\subsection{CAMIRA Cluster Catalog}
For this cluster WL analysis, we make use of the HSC CAMIRA cluster catalog \citep{Oguri2017}, based on HSC S16A data. The detailed methodology of the CAMIRA algorithm was presented in \cite{Oguri2014}. In short, it fits all photometric galaxies with the  stellar population synthesis (SPS) models of \cite{BC03} to compute likelihoods of being red-sequence galaxies as a function of redshift. From the likelihoods, a three-dimensional richness map is constructed to locate cluster candidates from peaks. For each cluster candidate, the brightest cluster galaxy (BCG) is identified, and in an iterative process the  richness, cluster photo-z, and the BCG identifications are then refined.

The CAMIRA-HSC catalog contains 1921 clusters at estimated redshifts $0.1 < z < 1.1$ and richness $\ncor> 15$, where richness is defined as the effective number of members above stellar mass greater than $10^{10.2}\ M_\odot$,
which roughly corresponds to $0.2L_*$. To be  conservative,  we only make use of clusters with  richness $\ncor> 20$ whose centers are within the FDFC area, totaling 921 clusters. The cluster redshifts are based on photo-z's (from the SPS model fitting described above) of the cluster member galaxies and show overall good performance, with  bias and scatter in $\Delta z/(1 + z)$ being better than 0.005 and 0.01 over most of the redshift range, respectively.

\section{Source Selection Methods}
\label{sec:SS}

\begin{table}
\tbl{Source sample properties for different selection methods}{
{
\begin{tabular}{ccccc}
\hline\hline
Sample                &     $n_g^a$ &  $N_s^b$& $\langle z_{\rm s,spec}\rangle^c$  & $\langle z_{\rm s,photo}\rangle^d$ \\
\hline
WL                       & 24.72 & $-$ & $-$ & $-$ \\
photo-z (`All')       & 21.66 &12 534 922 & 1.35 & 1.31 \\ 
CC-cut, $z_l<0.4$& 11.58 &5 049 427 & 1.24 & 1.25 \\
P-cut, $z_l<0.4$   & 13.80 &6 307 134 & 1.19 & 1.17 \\
CC-cut, $z_l\geq0.4$&7.05& 968 243 &  1.38 & 1.56 \\
P-cut, $z_l\geq0.4$  & 5.37& 999 337 &  1.46 & 1.54 \\
\hline
\end{tabular}
}
}
\label{tab:ndens}
\begin{tabnote}
$^b$ Number density (unweighted) in [arcmin$^{-1}$].\\
$^a$ Total number of source galaxies within 5~\mpch~ of the CAMIRA cluster centers.\\
$^c$ Mean source redshift, estimated from the lensing-weighted re-weighted spec-z's (see Section~\ref{subsec:spec-z}, Eq.~\ref{eq:lensreweight}).\\
$^d$ Mean source redshift, estimated from the lensing-weighted stacked $P(z)$ (see Section~\ref{subsec:spec-z}, Eq.~\ref{eq:stackedPz}).
\end{tabnote}
\end{table}

\subsection{Basic WL and photo-z cuts}
\label{subsec:SSall}
The WL shape catalogs used here contain galaxies that are within the footprint of the Wide survey and have well-measured shapes and photometry. To accomplish this, basic WL cuts have been explored and applied  to the galaxy catalogs and  are described in full detail in \cite{Mandelbaum2017}, and so we only briefly summarize them here. These include using galaxies within the FDFC area, excluding regions in which the PSF is poorly measured, and applying bright star masks \citep{Coupon2017}.  Photometry flags have been applied to remove objects with large deblendedness parameter, saturation, bad pixels, interpolated pixels, cosmic-rays, bad centroids, and those that have duplicate entries. Shape flags have been applied to keep galaxies with  ellipticities in the range $|e|<2$,  shape uncertainty in the range $0< \sigma_e<0.4$ and resolution factor $R_2 \geq 0.3$ \citep[as defined by the pipeline; see][]{Mandelbaum2017}. Finally, galaxies are limited to be brighter than  $i<24.5$, have $S/N>10$ in the $i$-band, and to have detections in at least two other bands.

Since photo-z's are used in the WL analysis, we further apply photo-z quality cuts on the WL catalog to remove undefined or inadequate photo-z's. In this paper we compare results from two photo-z codes: {\sc mlz} and {\sc franken-z}, though there are several others \citep{Tanaka2017}. In the case of {\sc mlz}, we require that the PDF standard deviation to be small, $\sigma(P(z))<3$, and have high photo-z ``confidence" factor, zConf$>0.13$ \citep[see ][]{Tanaka2017}. In the case of {\sc franken-z}, we require the  photo-z $\chi^2$ fit  to be small, $<6$. {In Section~\ref{sec:WLprof} we compare  WL profiles from these two photo-z methods, whereas elsewhere we adopt {\sc mlz} as the fiducial photo-z code.}

We combine the photometry, shape, photo-z and $P(z)$ information, applying the above quality cuts, to comprise a basic source catalog we refer to as `all'. The typical mean unweighted galaxy number density in the catalog is $21.7~{\rm arcmin}^{-2}$, depending on seeing \citep[see also Figure 9 in][]{Mandelbaum2017}. We summarize the mean number density  in {the first column of Table~\ref{tab:ndens} after applying the WL/photo-z cuts (first two rows). We also give the mean source redshift of the 'all' sample based on the redshift distribution that is estimated below in Section~\ref{subsec:spec-z}. The redshift distribution is calculated using two methods -- reweighted spec-z's (third column) and stacked $P(z)$ information (fourth column).}
As noted in Section~\ref{sec:WLmethod}, we make use of the full photo-z $P(z)$ when calculating the WL quantities, $\DSigma(R)$ and its weight, in a way that corrects for the dilution by objects lying in front of the lens, assuming that the $P(z)$ gives the true description of the galaxy redshift distribution. The further restrictive selection methods explored in subsequent subsections will explore  the validity of this assumption.

\subsection{Color-Color cuts}
\label{subsec:CC}

\begin{figure}
\includegraphics[width=0.5\textwidth]{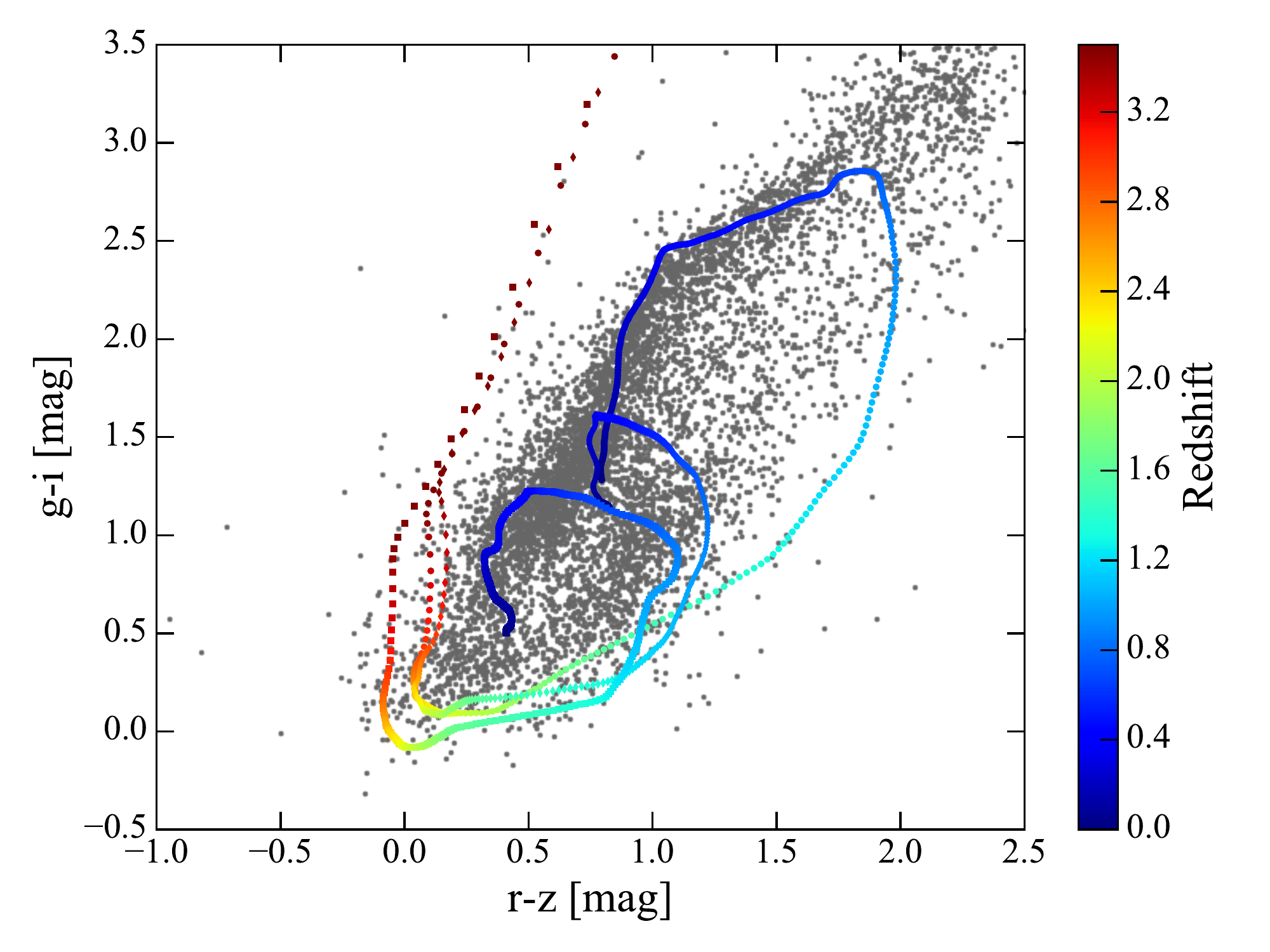}
\caption{$g-i$ versus $r-z$ CC diagram for galaxies in the HSC footprint. Black dots represent the color distribution of galaxies close ($R<100~\kpch$) to cluster centers, in order to highlight the red-sequence population. Colored tracks show the color evolution with redshift of synthetic galaxy templates of E, Sa and Sd-type galaxies (top right to bottom left). }
\label{fig:CCevtracks}
\end{figure}

\begin{figure*}
\includegraphics[width=0.5\textwidth]{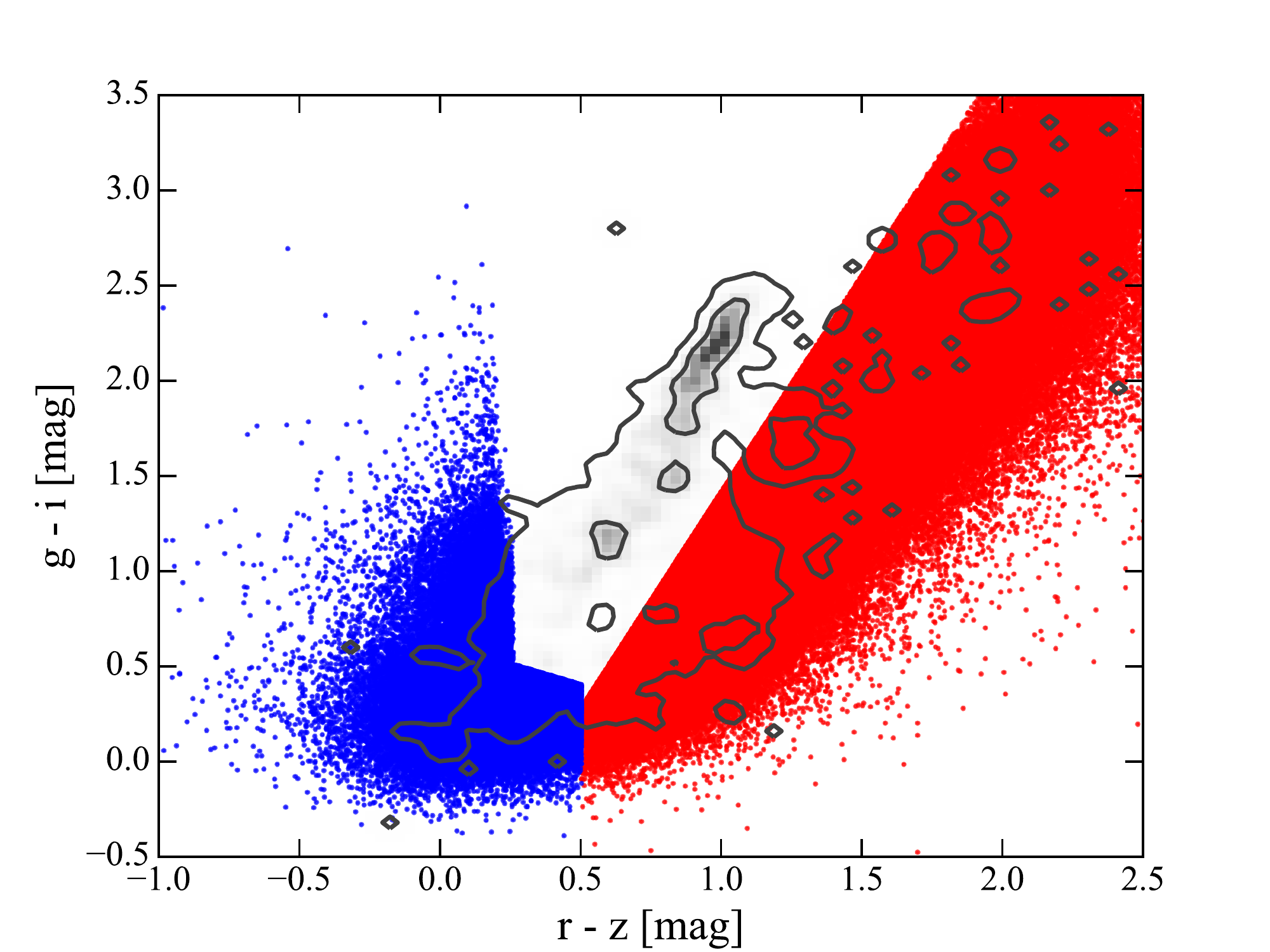}
\includegraphics[width=0.5\textwidth]{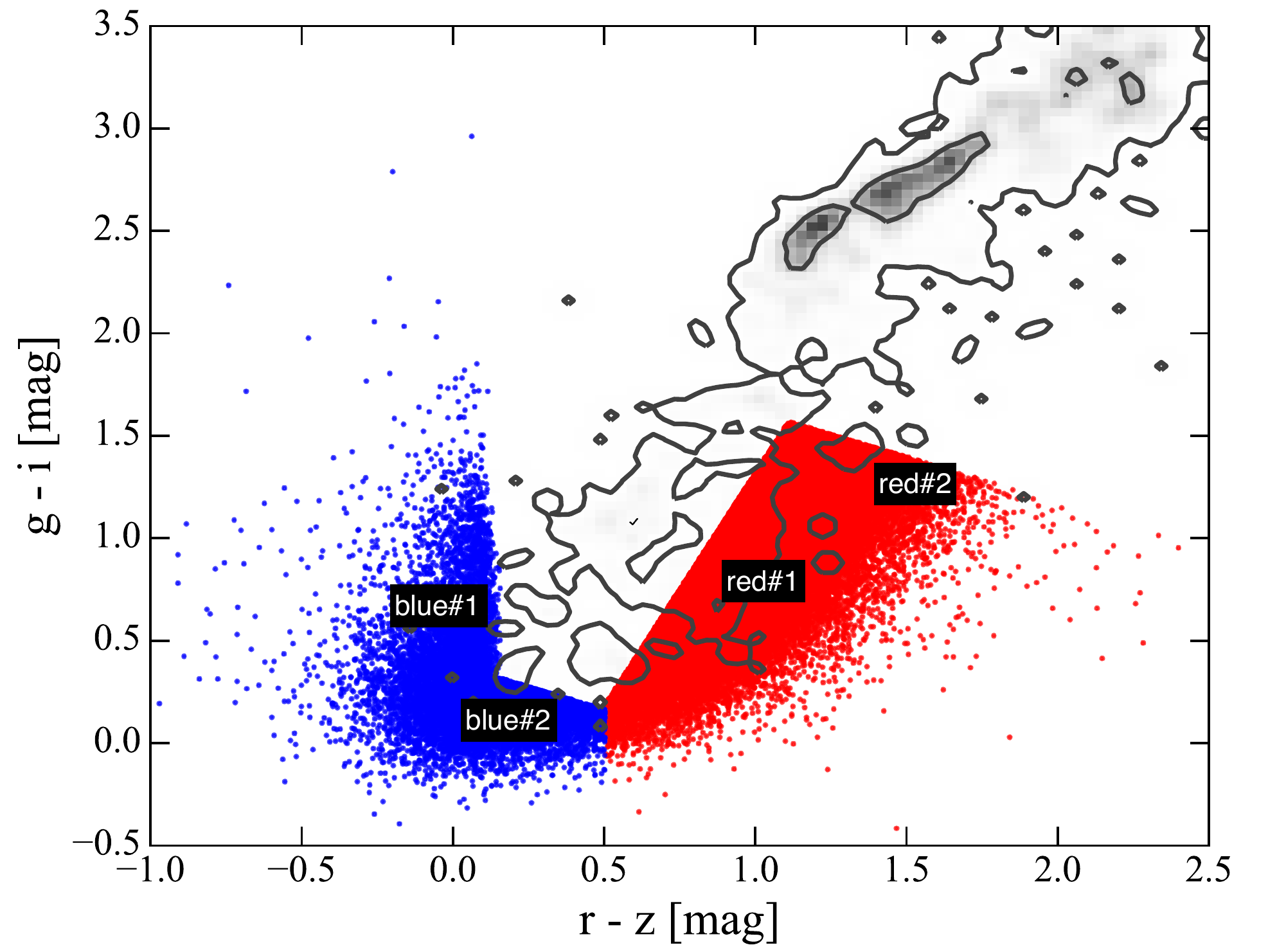}
\caption{Color-Color diagrams for galaxies in the HSC footprint. Contours and gray-scale colors show the distribution of galaxies close ($R<50~\kpch$) to cluster centers, in order to highlight the red-sequence population. Red and blue points denote the red and blue source populations, as selected by our fiducial CC cuts (denoted on the right panel as red\#1, red\#2, blue\#1, blue\#2, respectively). The left panel shows galaxies around $z<0.4$ clusters and the right panel shows galaxies around  $z\geq0.4$ clusters.}
\label{fig:CC}
\end{figure*}

\begin{figure*}[tb]
\includegraphics[width=1.\textwidth,clip]{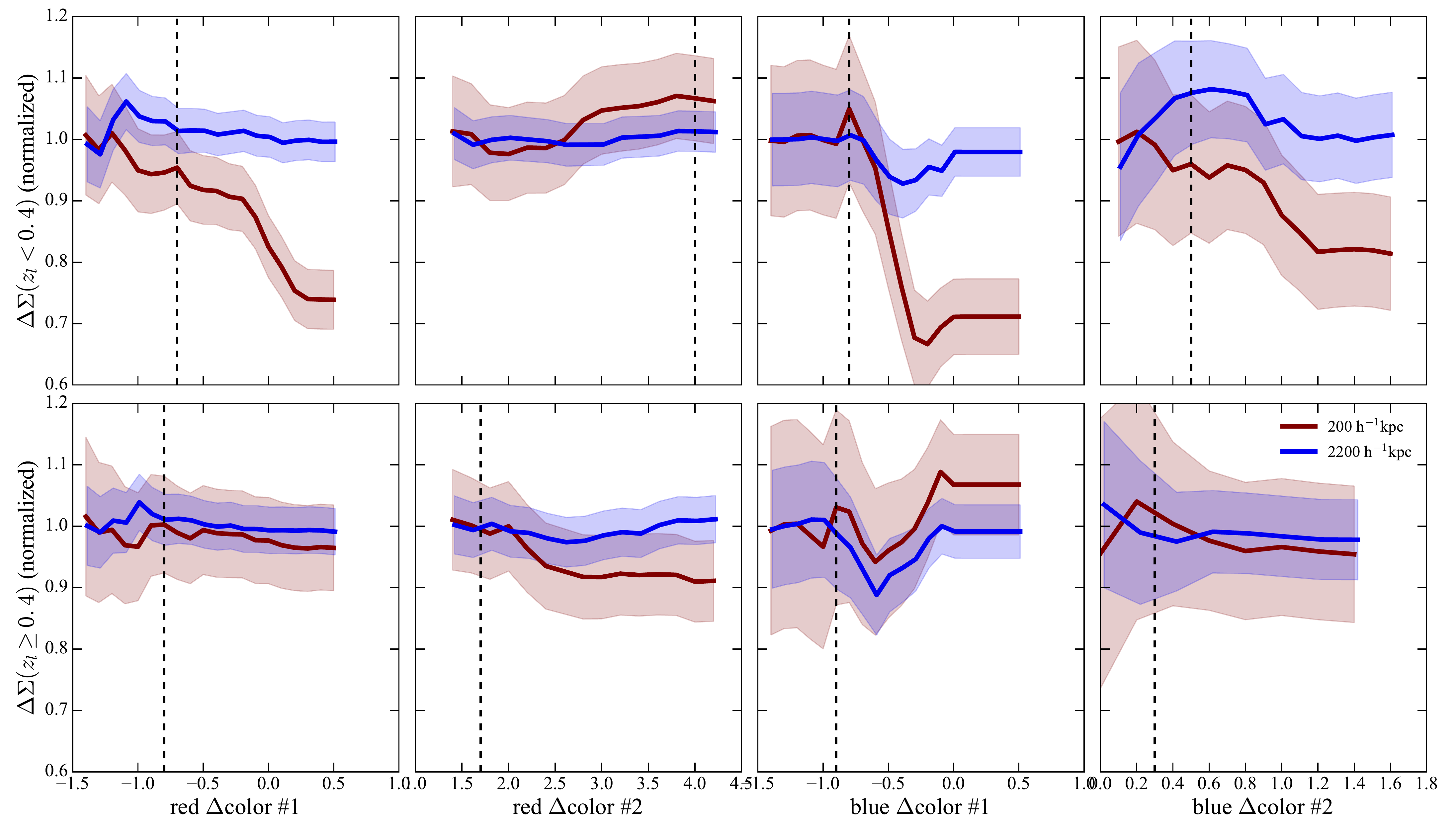}
\caption{Mean lensing signal, $\Delta\Sigma$  (normalized by the signal in the first color bin), as a function of each color limit of the CC-cut method, measured at two radial bins (colored by distance, indicated in the legends). Upper panels are for lenses at $z<0.4$, bottom panels are for lenses at $z>0.4$. From left to right, panels are for the red sample color limit \#1, the red sample color limit \#2, the blue sample color limit \#1, and blue sample color limit \#2, respectively, as indicated in the right panel of Figure~\ref{fig:CC}. Each color limit is a suitable combination of $g,r,i,z$  (see Appendix~\ref{app:CC}). In all panels, any contaminating population (e.g., red-sequence in red \#1 and \#2; foreground and/or blue cluster members in blue \#1 and \#2) are expected to lie to the right (i.e., larger $\Delta$color). As can be seen in some of the panels (e.g., red-\#1 and blue-\#1), the larger the color limit (advancing into the contaminating population), the smaller the mean lensing signal due to increased dilution.   The fiducial cuts selected for each color limits are indicated as vertical dashed lines. We note that the errors are highly correlated from left to right, as some of the same galaxies are used in the measurement. }
\label{fig:gt_colorlim}
\end{figure*}

\begin{figure*}[tb]
\includegraphics[width=1.\textwidth,clip]{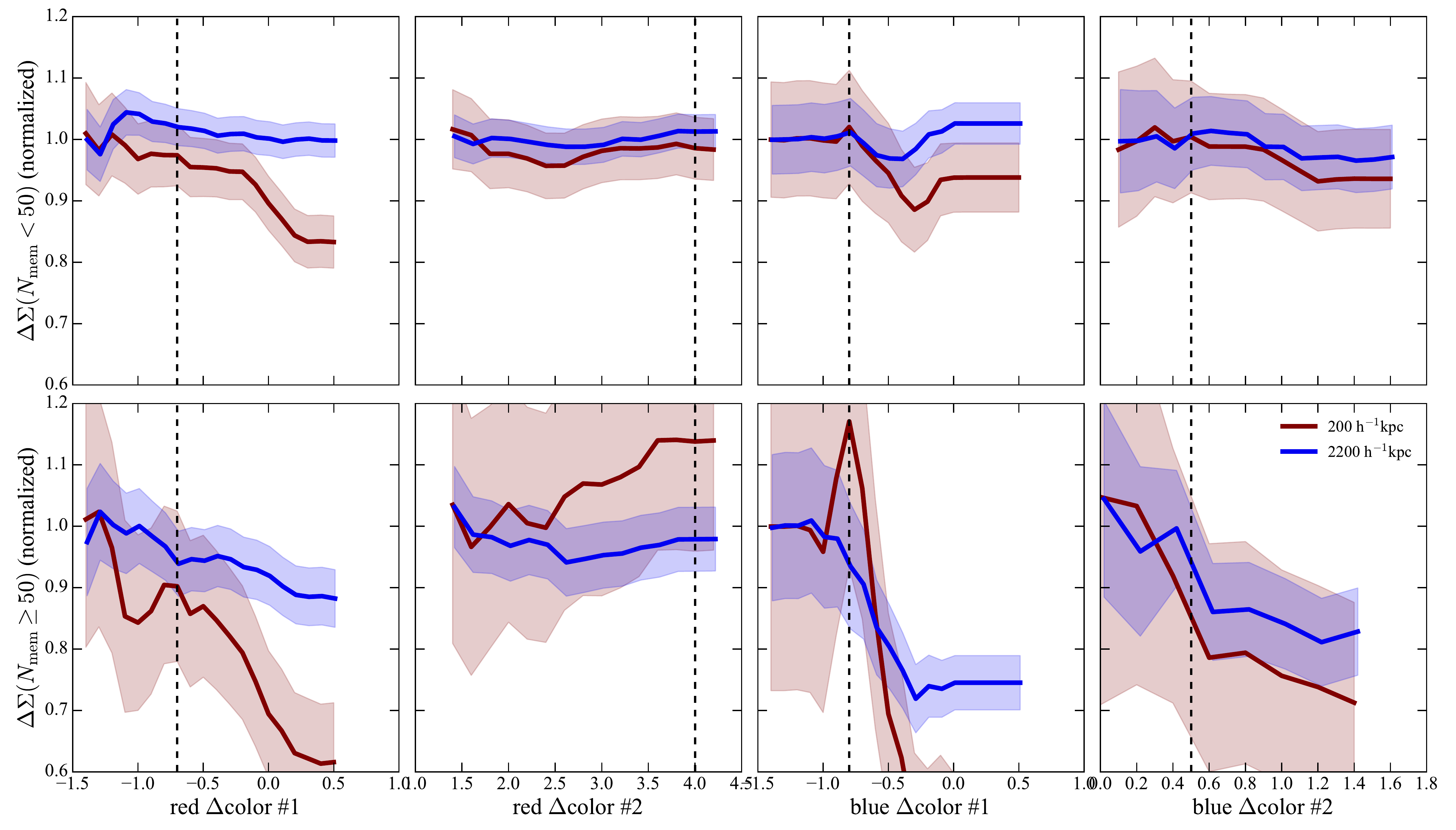}
\caption{Same as Figure~\ref{fig:gt_colorlim}, but divided into low richness ($\ncor<50$, top) and high richness ($\ncor\geq50$, bottom) clusters, instead of by cluster redshift. The vertical dashed lines denote the same fiducial cuts adopted for low-z clusters in Figure~\ref{fig:gt_colorlim}.}
\label{fig:gt_colorlim_N}
\end{figure*}

The color-color (CC) selection method has been extensively presented and explored in \cite{Medezinski2010} and used in the literature \citep{Medezinski2011,Medezinski2013,Medezinski2016,Oguri2012,Umetsu2012,Umetsu2014,Umetsu2015,Formicola2016,Wegner2017,Monteiro-Oliveira2017}. It is based on conservatively rejecting galaxies in areas of color space suspected as having a large fraction of cluster or foreground galaxies. The region of cluster members is easily identified in CC space by having an overdensity of red-sequence galaxies, particularly when plotting the number density in CC space of galaxies lying close to cluster centers; or alternatively, plotting the mean cluster centric distance in CC space (see figures 1 \& 2 in \citealt{Medezinski2010}). Using the HSC CAMIRA cluster catalog \citep{Oguri2017}, we further demonstrate it here. For HSC, we opt to use the $g-i$ vs $r-z$ CC space, which best spans the optical range given the bands observed in HSC and therefore maximizes the separation of different populations of galaxies in this space. 

To demonstrate this, similar to \cite{Medezinski2010},  we overlay on Figure~\ref{fig:CCevtracks} synthetic evolutionary tracks derived using {\sc galev} \citep{Kotulla2009} for different types of galaxies (E, Sa and Sd, top-right to lower-left colored curves) on the galaxy distribution (within $100~\kpch~$ of cluster centers; gray points) in $g-i$ and $r-z$ space.  This depicts where different populations are expected to lie. At lower redshifts the tracks lie close to $g-i\sim1,\ r-z\sim0.5$ where an overdensity of galaxies is seen. At these colors, galaxies  at $z\sim3$ drop out of the $g$ band (redder color of the curves at $r-z\sim0$ that extend redder in $g-i$ from 0--3.5), indicating a region of expected degeneracies between  low-z  ($\sim0.2$) $4000~\AA$-break galaxies and high-z ($\sim3$)  star-forming Lyman-break galaxies, causing large outlier fractions in photo-z's based on limited wide optical bands (especially due to the lack of $u$-band or deep IR to distinguish between the two). Thus we expect this region to contain a large fraction of foreground galaxies with possibly biased photo-z's.

We further select galaxies within $50~\kpch~$ from cluster centers in two cluster redshift bins, and plot their mean number density in the $g-i$ versus $r-z$  space in Figure~\ref{fig:CC} (gray-scale and contours; left for $z_l<0.4$ clusters, right for $z_l\geq0.4$ clusters). An overdensity of red-sequence cluster members is evident at $g-i\sim2.3,\ r-z\sim1$ for $z_l<0.4$ (left), and at $g-i\sim2.7,\ r-z\sim1.5$ for $z_l\geq0.4$ (right). To avoid  dilution of the lensing signal by cluster members, it is therefore important to exclude galaxies in this region from our source sample. 

To further explore the exact limits that best isolate background galaxies from the foreground and cluster regions, we hereby make use of the CAMIRA cluster catalog, and select background galaxies in two regions in CC space, marked by red and blue points in Figure~\ref{fig:CC}. Using this large statistical lens-source sample, we can demonstrate the effect of dilution by varying the color limits as we approach the suspected contaminating regions. We define several limits,  denoted  red-\#1 and red-\#2 for the red sample, which are approximately parallel and perpendicular to the cluster red-sequence and blue tail, and similarly blue-\#1 and blue-\#2 for the blue sample. These limits are marked on the right panel  of Figure~\ref{fig:CC}. The full description of the color limits may be found in  Appendix~\ref{app:CC}. In Figure~\ref{fig:gt_colorlim}, each panel presents the mean lensing signal, $\langle\DSigma\rangle$ (normalized by the leftmost 3 color bins) for each color limit, as we extend the limit further into the contaminating region  -- left to right panels are for limits red-\#1 and red-\#2 for the red source sample, and limits blue-\#1 and blue-\#2 for the blue source sample. They are further estimated separately for two cluster-centric annuli (brown and blue  curves as indicated in the legend), and for two lens redshift bins -- $z_l<0.4$ (upper panels) and $z_l\geq0.4$ (lower panels). In general, as we extend the color cut (to the right of each plot), we are approaching the contaminating population, and therefore expect the signal to drop due to dilution; however, we also increase the sample size by including more galaxies and therefore expect the shot noise to decrease, as indicated by the smaller uncertainties (shaded regions for each curve), scaling by the square-root of source number density. We note that cluster dilution will manifest as a decreasing  signal as a function of color limit for the inner radial bins only (brown curves), whereas dilution by foreground galaxies (of biased photo-z's of background galaxies) will manifest as decreasing signal at all radii, albeit inner radial bins will be noisier (due to lower statistical power). In all panels, the dashed vertical line marks the chosen cut below which sources are selected to have minimal contamination ($\lesssim 10\%$ relative signal dilution). We note, however, that the exact location of the cut to within about $\pm0.1$~dex does not significantly affect our results.

In  Figure~\ref{fig:gt_colorlim}, the panels that explore the mean lensing signal as a function of color limit red-\#1 and blue-\#1 behind lower redshift clusters (upper left and third from left panels, respectively) show clear signs of dilution, with  $\gtrsim20\%$ signal dilution in the inner radial bin (brown curve), but  $\lesssim10\%$ in the outer radial bins (blue curve), indicating mostly cluster contamination. 
The panel exploring the color limit blue-\#2 behind $z_l<0.4$ clusters (upper right) shows a $\sim20\%$ decrease in the inner radial bin, and a noisy and smaller ($\lesssim10\%$) decreasing trend in the outer radial bin, which may indicate foreground contamination. In the lower panels, for sources behind $z_l\geq0.4$ lenses, the trends are not as significant, except perhaps for red-\#2 and blue-\#2, where residual cluster dilution may remain but at the $<10\%$ level. This is probably  because the effective source density is lower, leading to noisier trends.  
It may be that the cluster contamination is less severe at higher lens redshifts simply because fewer faint members are detected. As is also evident from Figure~\ref{fig:CC} (right),  high-z clusters occupy a redder region in CC space that is  better separated from the background red and blue samples.
To be conservative, we  set the limits as indicated by the vertical dashed lines. We display the final CC selection of red and blue (red and blue points) background galaxies in Figure~\ref{fig:CC} behind $z_l<0.4$ and $z_l\geq0.4$ clusters (left and right, respectively). We summarize the unweighted galaxy number density {and mean source redshift} after applying these sets of cuts in Table~\ref{tab:ndens} (third and fifth {rows}) for a simple comparison. As can be seen, for the low-$z_l$ cuts, this selection removes about 50\% of the galaxies, whereas at high-$z_l$, it removes about 67\% of the objects.

Finally, in Figure~\ref{fig:gt_colorlim_N}, we carry out a similar exercise,  exploring the mean lensing signal as a function of color cut, now separating into two cluster richness regimes, low ($\ncor<50$, top) and high ($\ncor\geq50$, bottom). The dilution by cluster members is now  stronger for high richness clusters, as expected. Within the noise, it appears that our chosen CC for the low-$z_l$ regime still succeed in removing most of the contamination for both richness bins.

\subsection{$P(z)$ cuts}
\label{subsec:Pcut}

\begin{figure}[tb]
\includegraphics[width=0.5\textwidth,clip]{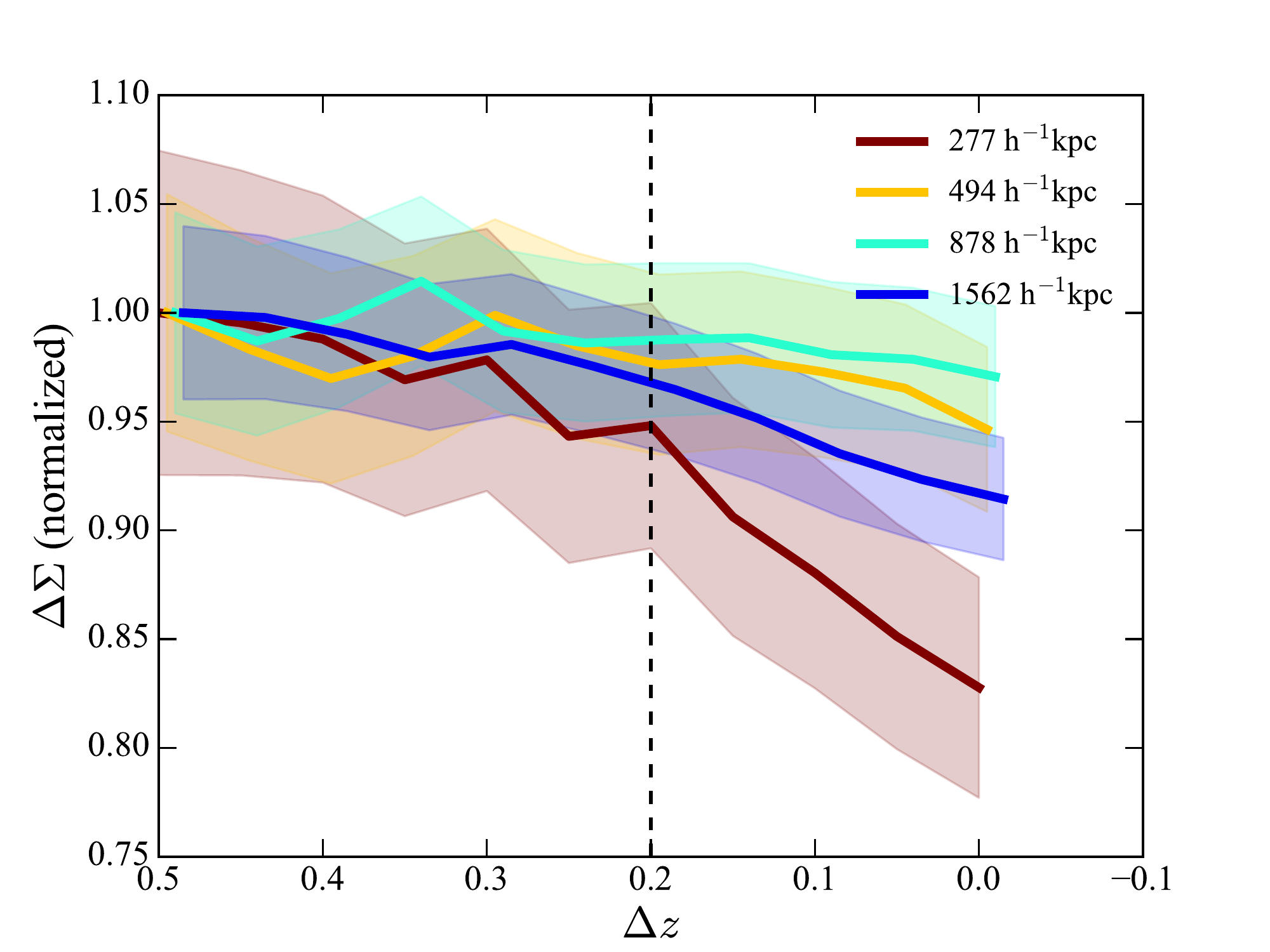}
\caption{Mean lensing signal, $\Delta\Sigma$ (normalized by the signal in the largest $\Delta z$ bin), as function of lens redshift  threshold, $\Delta z$, defined in Equation~\ref{eq:Pcut} for the P-cut method, measured at several radial bins (colored by distance).  The vertical dashed line indicates the optimal redshift threshold, $\Delta z=0.2$, that minimizes contamination. }
\label{fig:gt_dzlim}
\end{figure}

\begin{figure*}[tb]
\includegraphics[width=\textwidth,clip]{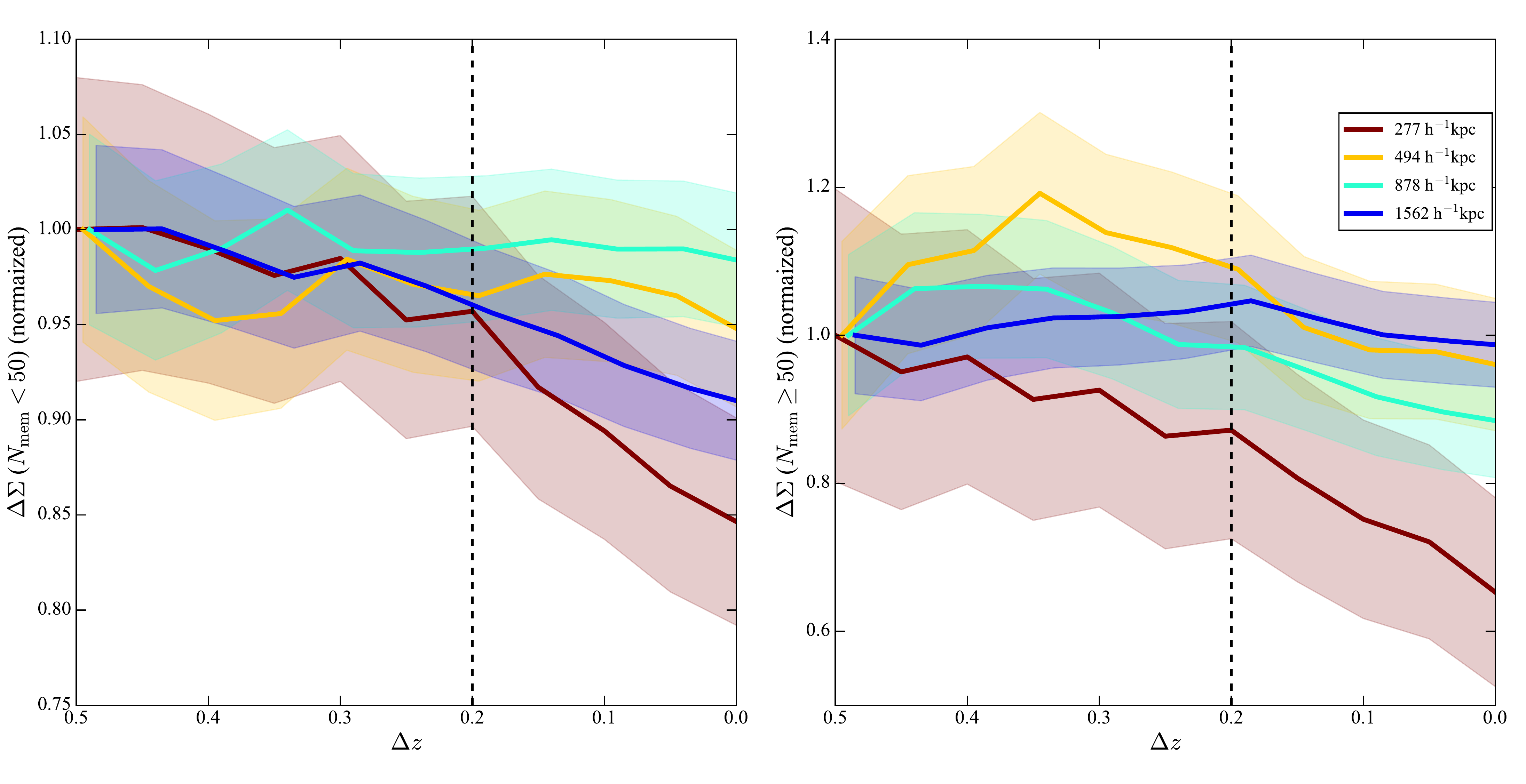}
\caption{Same as Figure~\ref{fig:gt_dzlim}, further divided into low richness ($\ncor<50$, left) and high richness ($\ncor>50$, right) clusters. For the high richness bin, cluster dilution is present even for the highest threshold, $\Delta z=0.5$. }
\label{fig:gt_dzlim_N}
\end{figure*}

A second secure source selection method examined in this paper has been presented in \cite{Oguri2014}. It relies on photo-z selection, but rather than using photo-z point estimates, this approach utilizes the full $P(z)$ information for each galaxy (hereafter P-cut). With this method, we define a sample of galaxies that satisfy:
\begin{equation}\label{eq:Pcut}
\int\limits_{z_{l}+\Delta z }^{\infty} P(z)\,\mathrm{d}z >p_{\rm cut} \textrm{ and } z_{p}<z_{\rm max}
\end{equation}
where $P(z)$ is the photo-z PDF and $z_p$ is the redshift point estimate. Here we choose to use the Monte-Carlo derived point estimate, \texttt{photoz\_mc}. In \cite{Oguri2014}, the  minimum redshift cut used was $z_l+0.05$, i.e. a threshold of $\Delta z=0.05$ above the cluster redshift. The sample was further defined such that   $p_{\rm cut}=0.98$, i.e., 98\% of the $P(z)$ lies beyond this lens redshift threshold. Finally, the maximum redshift was set  to $z_{\rm max}=1.3$, since for the CFHTLenS data used in that WL analysis, the photo-z's above that limit  are thought to be less secure \citep{Kilbinger2013}.

In this section we further attempt to establish the required cuts more robustly, by exploring them in a similar fashion as introduced in the previous section, using the sources behind CAMIRA clusters in  HSC data. First, since HSC data are much deeper than the CFHTLenS sample used in \cite{Oguri2014}, and extend to the $y$-band, we  set $z_{\rm max}=2.5$. We also adopt as before, $p_{\rm cut}=0.98$. We vary both of these limits, in the range $z_{\rm max}=1.5$--$3$ and $p_{\rm cut}=0.97$--$1$, and measure the mean lensing signal, but find this has little to no effect on the recovered lensing signal. On the other hand, we find that varying $\Delta z$ has a significant dilution effect. In Figure~\ref{fig:gt_dzlim}, we present the mean lensing signal, $\langle\DSigma\rangle$ (normalized), as a function of $\Delta z$ for all lens-source pairs, and measured in different cluster-centric annuli (color-coded in the legend). As can be seen, up to $\Delta z=0.2$, there is a significant dilution of the lensing signal. It is most significant in the innermost radial bin (brown curve), suggesting that it is due to contamination by cluster members. 
However, we note that  the outermost radial bin (blue curve) also shows some decrease, indicating that some of the contamination is due to foreground galaxies or faint cluster members. In this selection scheme, it is not possible to separate cluster from foreground contamination as in the CC-cut scheme.
This plot overall indicates that  contamination is present up to a higher threshold than previously adopted ($\Delta z=0.05$).
We therefore set the limit to $\Delta z=0.2$ in  subsequent analyses.

Here we also repeat the test of dividing the sample into sources behind low- and high-richness clusters. Figure~\ref{fig:gt_dzlim_N} shows the mean lensing signal as a function of redshift threshold for two richness bins. The selected cut, $\Delta z=0.2$ still applies for low-richness clusters, but as can be seen from the right panel, for high-richness clusters the dilution in the innermost bin is present even at $\Delta z=0.5$, albeit the uncertainty on these curves is much larger due to the small number of clusters available for this test. This method therefore seems less successful in removing the cluster dilution for very massive clusters.

The unweighted galaxy number density {and mean source redshift} after applying the chosen cuts are summarized in Table~\ref{tab:ndens} (fourth and sixth {rows}). We set $\langle z_l\rangle=0.27$ for a selection relative to low-z clusters, and $\langle z_l\rangle=0.7$ for a selection relative to high-z clusters (since a lens redshift must be assumed for this procedure), to compare with the typical CC selection for $z_l<0.4$ and $z_l\geq0.4$. Similar to the CC selection, at low lens redshift about 35\% of galaxies are removed, and at high lens redshifts about 75\% are removed. 

\section{Results}
\label{sec:WLprof}


\begin{figure*}
\includegraphics[width=\textwidth,clip]{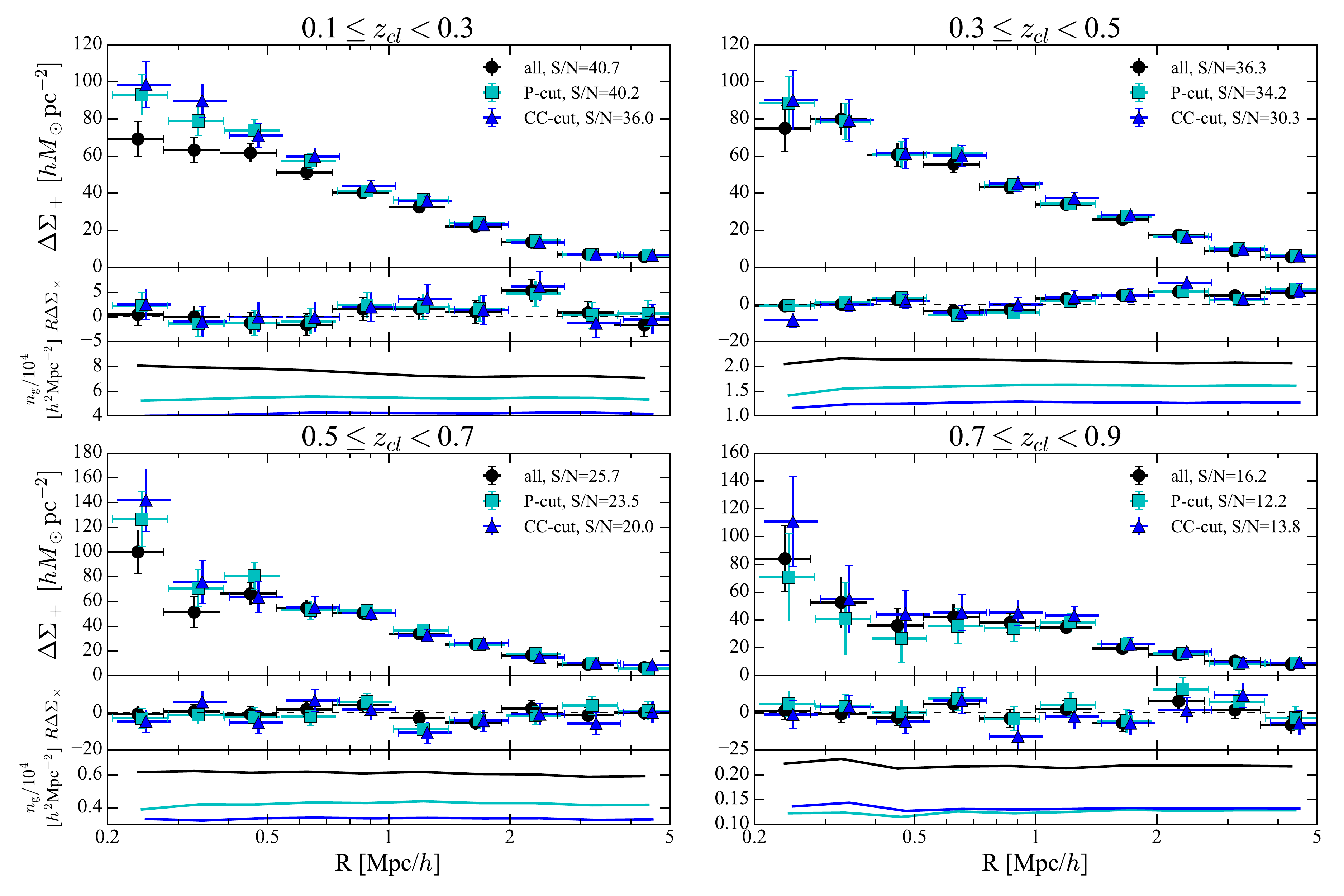}
\caption{Stacked WL profiles around CAMIRA clusters, comparing between different source selection methods, for four different lens redshift bins, as indicated. Top panels show the surface mass density contrast profile, $\DSigma$, middle panels show the $45\degree$-rotated shear (B-mode), expected to agree with zero; bottom panels show the effective source number density. All quantities in this plot were calculated using the {\sc mlz} photo-$z$ PDFs. Different lines in each panel show different source selection schemes: using all galaxies behind the lens and simply incorporating $P(z)$ (black), using P-cut selected galaxies (for which 98\% of $\sum P(z)$ lies behind $z_{l}+0.2$; cyan), or CC-cut selected galaxies as depicted in Table~\ref{fig:CC} (blue). S/N values for each selection are given in the legend of each panel.}
\label{fig:WLprof_zbins}
\end{figure*}
\begin{figure*}
\includegraphics[width=\textwidth,clip]{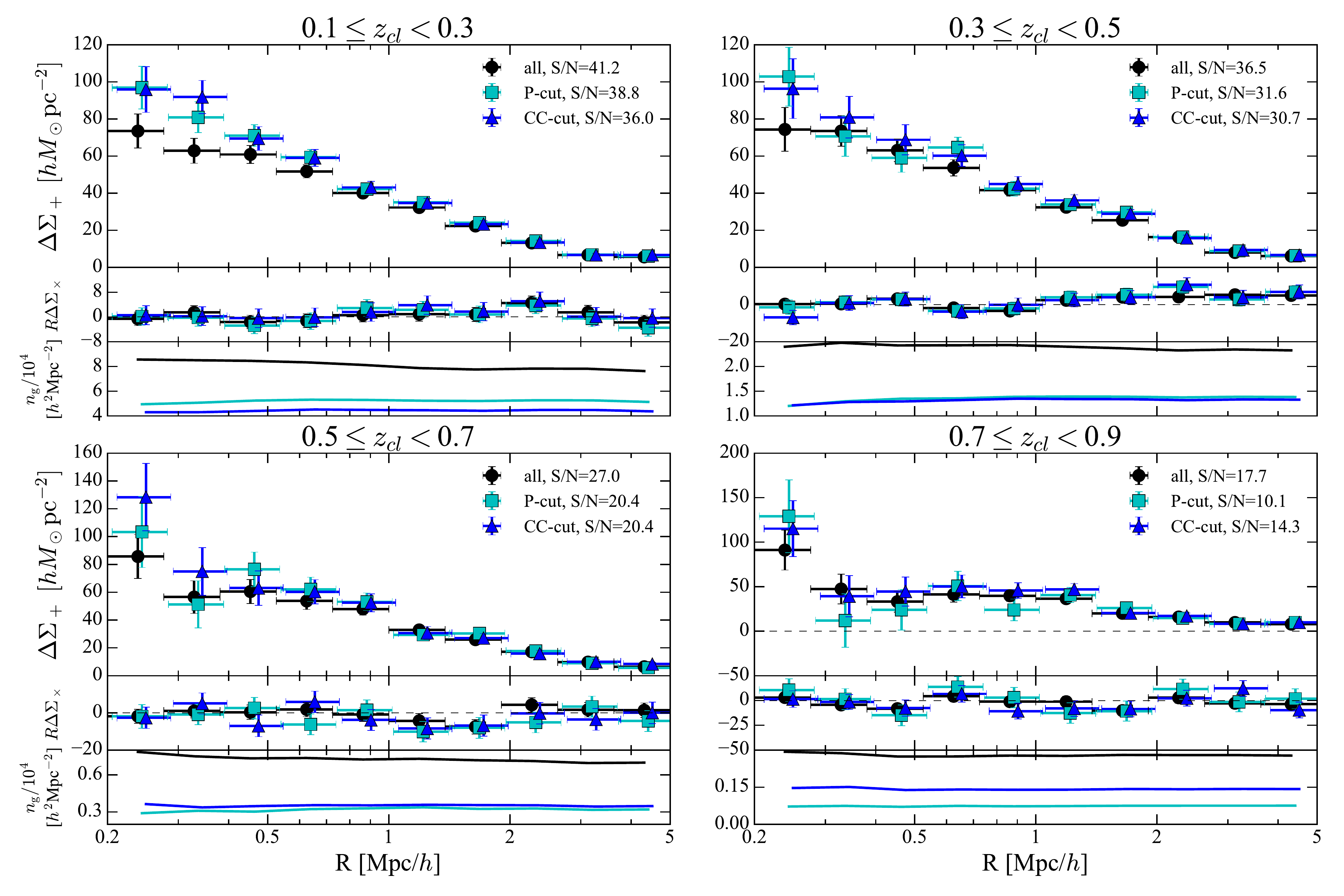}
\caption{Same as Figure~\ref{fig:WLprof_zbins}, but using the {\sc franken}-z photo-$z$ PDFs. }
\label{fig:WLprof_zbins_frankenz}
\end{figure*}

\begin{figure*}
\includegraphics[width=\textwidth,clip]{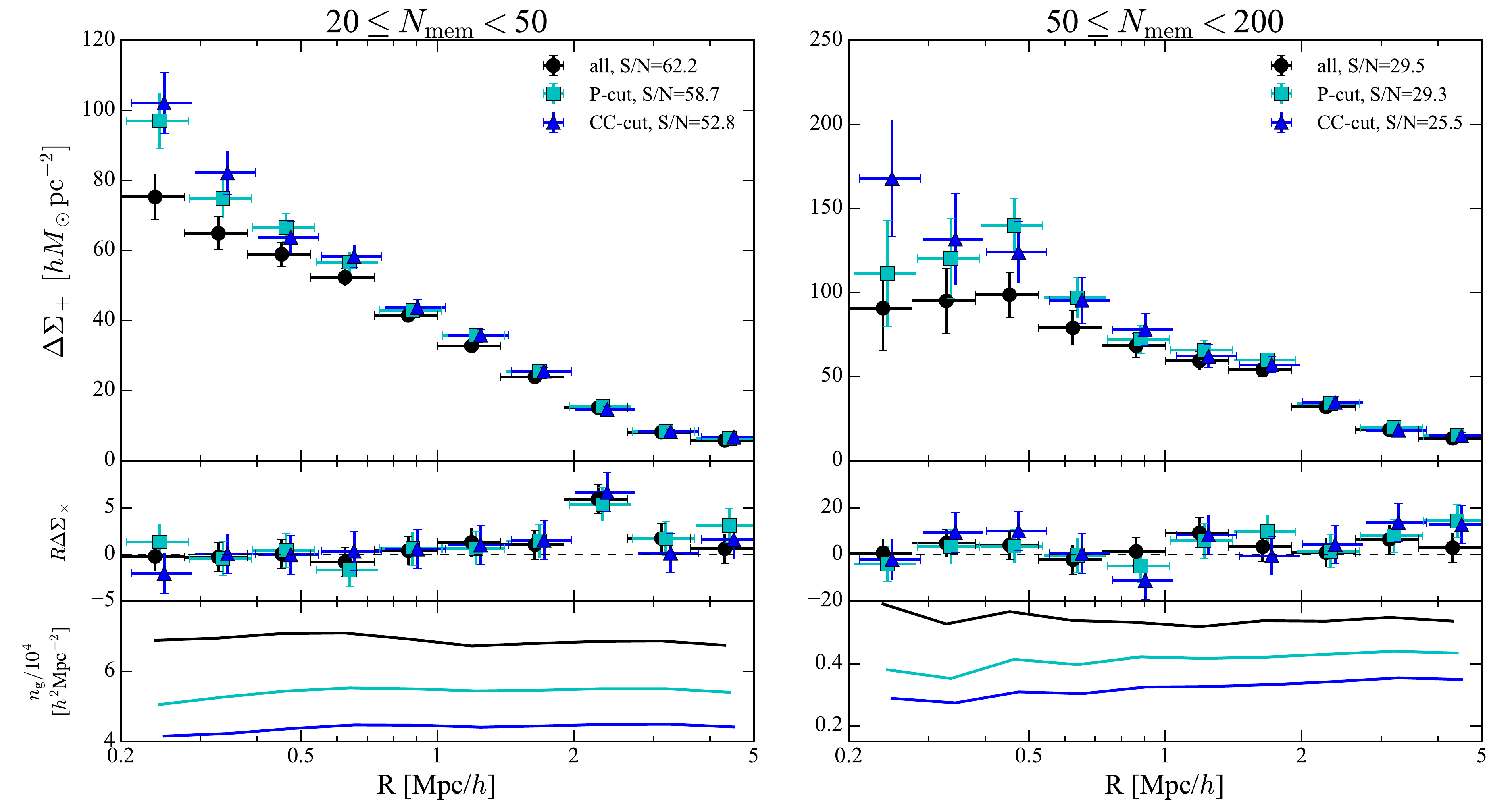}
\caption{Same as Figure~\ref{fig:WLprof_zbins}, but for different richness bins: left for $20\leq \ncor<50$ and right for $50\leq \ncor$. The CC and P-cut selections are consistent for low richness clusters, but for high richness clusters the P-cut method shows a hint of dilution in the innermost bin. }
\label{fig:WLprof_Nbins}
\end{figure*}

In this section, we present and compare the  cluster WL profiles, as derived using the different source selection methods described in the previous section. In the top panel of each subplot in Figure~\ref{fig:WLprof_zbins}, we calculate the surface mass over-density, $\DSigma(R)$, in 10 logarithmically-spaced bins spanning $0.2$--$5~\mpch$, for four different lens redshift bins spanning 0.1--0.9 (top left to bottom right). The errors represent the statistical uncertainties due to shape noise. The middle panel of each subplot shows the $45\degree$-rotated shear. As expected, this cross-shear is consistent with zero within the errors. Finally, the bottom panel of each subplot shows the {\it effective} source density profile, after accounting for the lensing weight (see Equation~\ref{eq:lensweight}; so this depicts the  density of only those sources used in the lensing calculation as determined by the photo-z PDF). The  different curves show different selection schemes -- using `all' galaxies (i.e., only WL+photo-z cuts applied and incorporating the photo-z PDF; black circles), after applying P-cut (cyan squares), and after applying CC-cuts (blue triangles). For all lens redshift slices, the CC-cuts (blue) provide profiles that are consistently higher than without the cuts (black), especially for the case of low-z lenses (upper left panel). When comparing CC-cuts and P-cut profiles, the conservative cut made for P-cut, $\Delta z=0.2$, results in very consistent profiles within the errors up to $z_l<0.7$. At the highest lens bin (bottom right),  P-cut gives a somewhat lower signal than the other methods, but consistent within the large errors. In conclusion, for nearly all lens redshifts, not applying any cut but rather relying on the photo-z PDF to correct for dilution will result in significantly diluted profiles (black).

To test how much these biases  are due to photo-z codes, we reproduce the same plots using the {\sc franken}-z photo-z code  (Speagle et al., in prep)  in Figure~\ref{fig:WLprof_zbins_frankenz}. Overall the same trend is observed, where for all lens redshift bins, CC-cuts and P-cut profiles agree, and slight differences are seen only for the highest lens bin (lower right panel). A more comprehensive comparison between the performance of different photo-z codes, of which seven different variants are run for HSC, will be discussed in More et al. (in prep), and so we defer this discussion there. 
We will note that  More et al.   also find that the differences in the redshift distributions derived from different photo-z codes are most apparent at higher redshifts, $z>1$. In this case, our high-$z_l$ lensing profile (lower right panel of Figure~\ref{fig:WLprof_zbins}) will also be most affected by photo-z code differences, since most of the sources in that bin lie beyond $z>0.7$.

The source density profiles in the bottom of each subplot show that the P-cut method provides more source galaxies at most lens redshift slices, up to $z_l\sim0.7$, where both P-cut and CC-cuts remove the same fraction of galaxies (for {\sc franken-z}, P-cut removes even more). The same is also indicated by the S/N level in the legend of each subplot and the raw number densities in Table~\ref{tab:ndens}. In conclusion, the P-cut method appears to perform slightly better for low-z clusters, and slightly worse for high-z clusters, than the CC selection. We also note that using `all' galaxies results in number density profiles that show somewhat of an excess at small scales compared to the conservative selection methods, especially noticeable at low redshifts (top left panel). This is, as discussed, due to cluster contamination, although the effect is harder to see in this plot since the profiles are not normalized. We will address and compare number density profiles more clearly below (see Section~\ref{subsec:boost}).

Since dilution by cluster members is expected to be worse for high richness clusters, where the fraction of cluster galaxies is by definition higher, we also explore the performance of these methods as a function of cluster richness. In Figure~\ref{fig:WLprof_Nbins} we plot the lensing profiles for each method in two richness bins, $20\leq \ncor\leq50$ (left) and  $50\leq \ncor$ (right). At low richness, as before, the CC- and P-cut methods show consistent profiles, whereas applying no cuts (`all') results in a lower signal profile. For high-richness clusters, on the other hand, the P-cut method appears slightly diluted at the innermost bin relative to the CC-cut profile, although they are consistent within the large statistical errors.  For this measurement,  only  38 high-richness clusters are available. When the full HSC-Wide survey is complete, this type of test can be done with smaller statistical errors, as we expect to have $\sim200$ $\ncor>50$ clusters in the full CAMIRA catalog.
We conclude from this test that extra care has to be taken in removing cluster contamination when studying very rich (or massive) clusters, in which case the CC selection method is then preferred.

\subsection{NFW Modelling}
\label{subsec:NFW}

\begin{figure*}[tb]
\includegraphics[width=0.56\textwidth,clip]{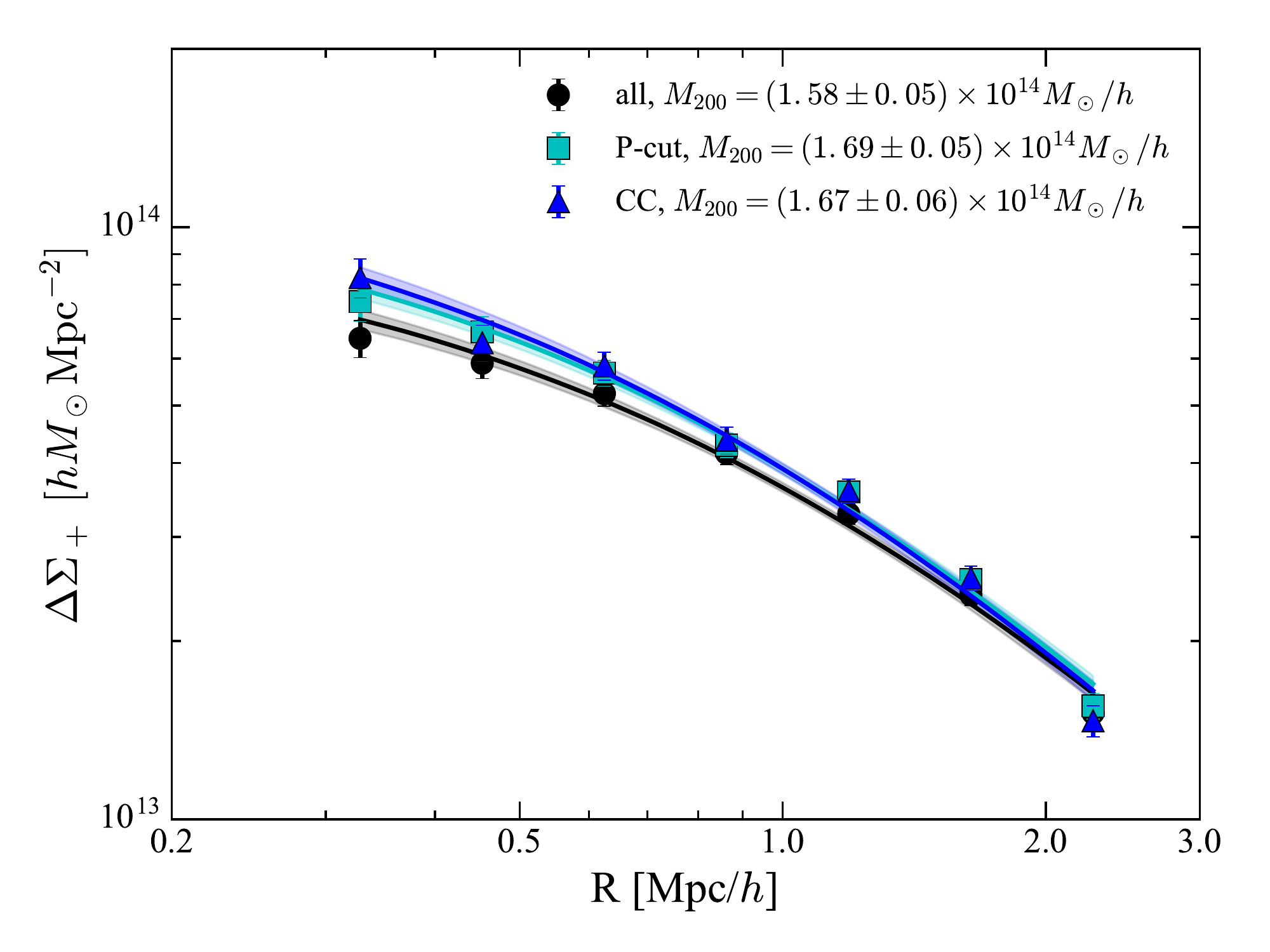}
\includegraphics[width=0.44\textwidth,clip]{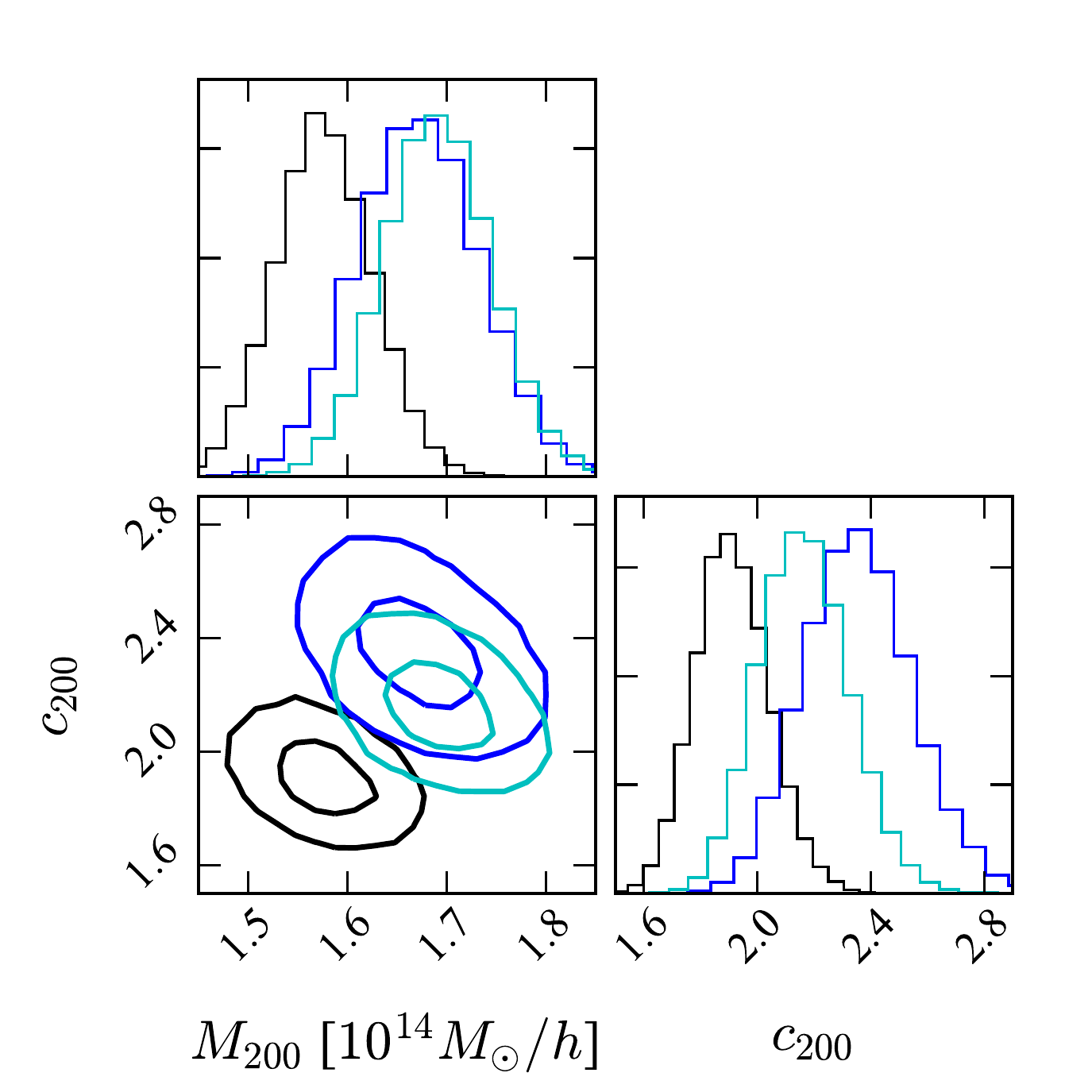}
\caption{NFW model fits to the WL profiles obtained from different selection selection schemes. Left: datapoints show the stacked surface mass density profiles, $\DSigma(R)$, for all galaxies (black), CC-cut galaxies (blue) and P-cut galaxies (cyan), and solid lines and shaded regions show the equivalent NFW fit. The median total mass from each fit is given in the legend. Right: 1,2-$\sigma$ confidence levels on the posterior 1- and 2-D distributions of the fitted NFW parameters, \mvir~ and \cvir, for each selection method (same color scheme as left panel).}
\label{fig:NFW}
\end{figure*}

\begin{table}
\tbl{Best-fit NFW model parameters}{
\begin{tabular}{ccccc}
\hline\hline
Method & $M_{\rm 200c}$ & $c_{\rm 200c}$ & $M_{\rm 500c}$ & $c_{\rm 500c}$   \\
& [\mhunit] && [\mhunit] & \\
\hline
\hline
All & $1.58\pm0.05$ & $1.91\pm0.13$ & $0.89\pm0.03$ & $1.15\pm0.08$ \\ 
P-cut & $1.69\pm0.05$ & $2.16\pm0.16$ & $1.01\pm0.03$ & $1.39\pm0.10$ \\
CC-cuts & $1.67\pm0.06$ & $2.35\pm0.19$  & $1.02\pm0.03$ & $1.52\pm0.12$ \\
\hline
\end{tabular}
}\label{tab:NFW}
\end{table}

The interesting quantity for deriving cosmological cluster counts is the total cluster  mass. Furthermore, the shape of the profile, as quantified by its concentration, provides insight into the formation history of a cluster \citep[e.g.,][]{Umetsu2014}, and $\Lambda$CDM simulations give predictions for this mass-concentration relation \citep{Bhattacharya2013,Dutton2014}. To derive the total mass and  concentration, and demonstrate the effect of source dilution on these quantities, we fit our stacked $\DSigma(R)$ profiles for each selection method with a spherically symmetric 
 central \citet*[NFW]{NFW96} model.  The free parameters in this model are the mass,  $M_{200c}$, and concentration,  $c_{200c}$, both in overdensity of 200 times the critical density of the universe.
  We fix the lensing-weighted mean cluster redshift and fit for the mass and concentration using the Markov Chain Monte Carlo (MCMC) algorithm {\sc emcee}  \citep{Foreman-Mackey2014}. We exclude the innermost radial bin, where masking  and deblending of BCGs may affect our photo-z's or shape measurements (see discussion in Section~\ref{subsec:boost}; also Murata et al. 2017).
 We also exclude the last two radial bins from the NFW analysis since we are only interested in the 1-halo term (the cluster) in this fit, and fit in the range $0.3$--$3$~\mpch. For the sake of computational efficiency, we set flat priors on the mass and concentration in the range $0\leq \mvir/\mhunit\leq100$, $0\leq \cvir\leq10$. 
 {We do not include miscentering in our model since it will equally affect all our source selection methods. Here we are only interested in the effect of dilution on our selections. We note, however, that neglecting the effect of miscentering will lead to overall lower concentrations, as compared with \lcdm~ predictions, and slightly lower masses.}
 {As mentioned in Section~\ref{sec:WLmethod},  the covariance only includes the statistical uncertainty due to shape noise, since we are only interested in comparing the systematic error due to source selection with the uncertainty induced by statistics. In what follows we have tested the effect covariance due to uncorrelated large-scale structure, and find it is negligible, since we are stacking over about a thousand clusters in a wide enough area. }

Here we fit the WL profile stacked over all CAMIRA clusters in the redshift range $0.1<z_l<1.1$ without subdividing into lens redshift or richness slices as in the previous section. The resulting profile (points) and its best-fit NFW profile (solid curves with shaded error interval) are shown in the left panel of Figure~\ref{fig:NFW} for each of the selection methods (`all' in black, P-cut in cyan, CC in blue). The corresponding posterior distributions of the mass and concentration fitted parameters from the MCMC chains are shown in the right panel of  Figure~\ref{fig:NFW}, with contours representing 1,2-$\sigma$ confidence bounds. The fitted values {and their statistical uncertainties} for each method are  summarized in Table~\ref{tab:NFW}. Since quantities derived by other mass proxies (e.g., X-ray, SZ) are often quoted in overdensities of $\Delta=500$, we also convert and quote  $M_{\rm 500c},\ c_{\rm 500c}$ in Table~\ref{tab:NFW}.
{To be complete, we furthermore fit in the same way the stacked lensing signal derived from the CC-cut method for clusters at low ($z_l<0.4$) and high ($z_l<0.4$) redshifts. We find mean masses of $\mvir = (1.52\pm0.06)\times\mhunit$ for low-z clusters, and $\mvir = (2.01\pm0.14)\times\mhunit$ for high-z clusters. These statistical mass constraints (4\% for low-z and 7\% for high-z) set the tolerance for the required systematic level.}

We now estimate how much bias is caused to the $M_{\rm 500c},\ c_{\rm 500c}$ derived values by dilution. In order to account for the statistical correlation between the `all' and CC-cut samples due to the latter being strictly a $\sim$50\% subset of the former, we bootstrap each of the source samples 100 times, and follow the same stacking and fitting procedure. We find that using `all' galaxies results in a mass that is underestimated by $1-M_{\rm 500c, all}/M_{\rm 500c, CC} = (13\pm 4)$\%. The level of bias on the concentration parameter is higher, and can cause an underestimation by as much as $1-c_{\rm 500c, all}/c_{\rm 500c, CC} = (24\pm11)\%$ when comparing between `all' and CC-selected sources. Although not highly significant here, this level of bias, if true, may emerge in future surveys such as LSST detecting thousands of clusters with  percent level statistical errors on the mean mass. 

\section{Validation Tests}
\label{sec:tests}

So far, we have relied on comparing the  lensing profiles of the different source selection methods as an indication that they minimize cluster and foreground contamination.  
Since the signal is a combination of both the shear and the redshift of the galaxies, 
all methods may still suffer from the same redshift bias that would not be apparent in such a relative comparison. {Furthermore, some residual cluster contamination may still be present in our CC- or P-cut samples.
To  more directly test the level of contamination, we present in this section independent validation tests. We first estimate the level of photo-z bias using  spec-z's in section~\ref{subsec:spec-z}, and in  section~\ref{subsec:boost} we examine the use of boost factors to estimate the residual cluster member contamination.}

\subsection{Photo-z systematics from re-weighted Spectroscopic Redshifts}
\label{subsec:spec-z}

\begin{figure*}[tb]
\includegraphics[width=0.5\textwidth,clip]{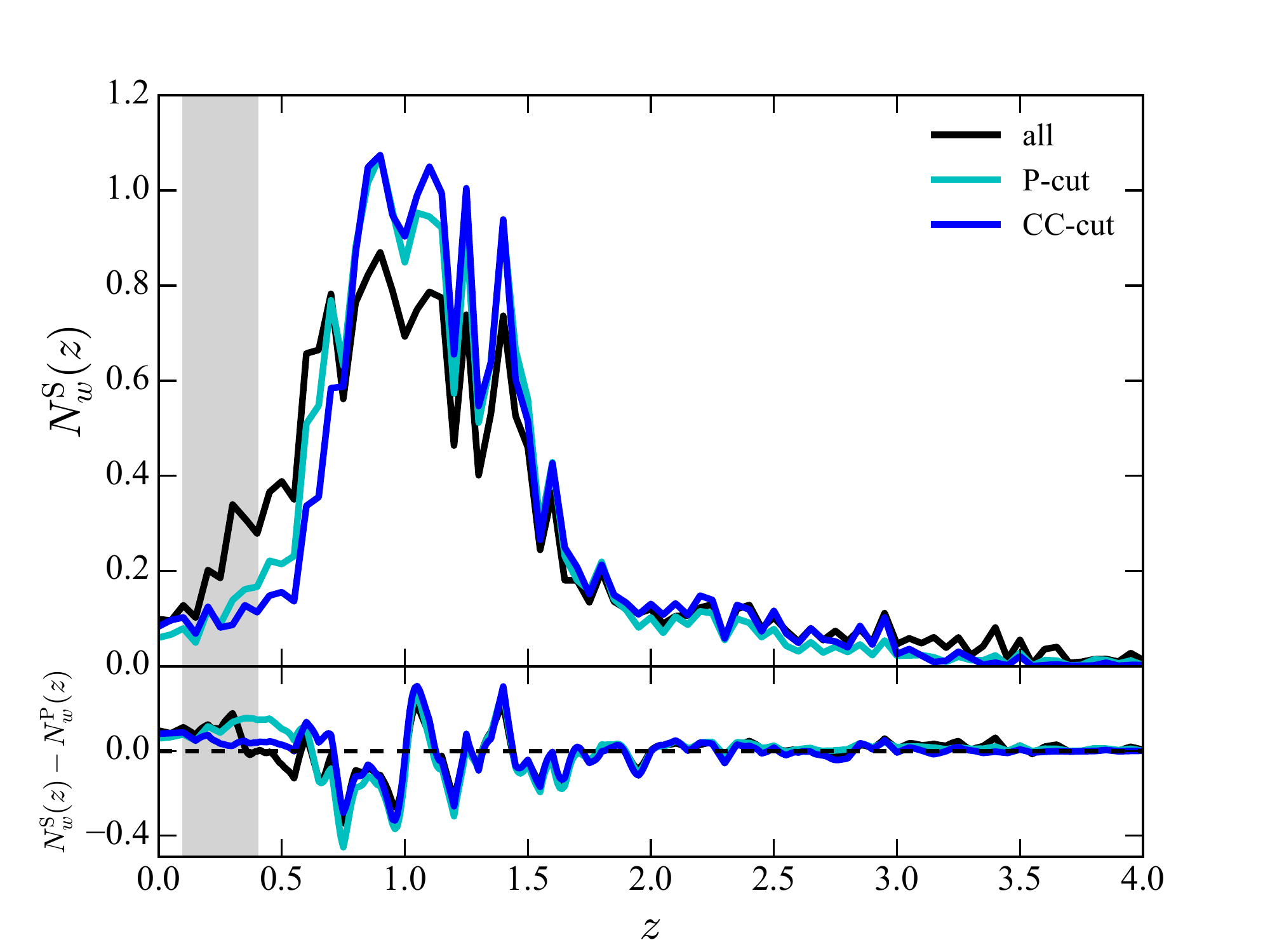}
\includegraphics[width=0.5\textwidth,clip]{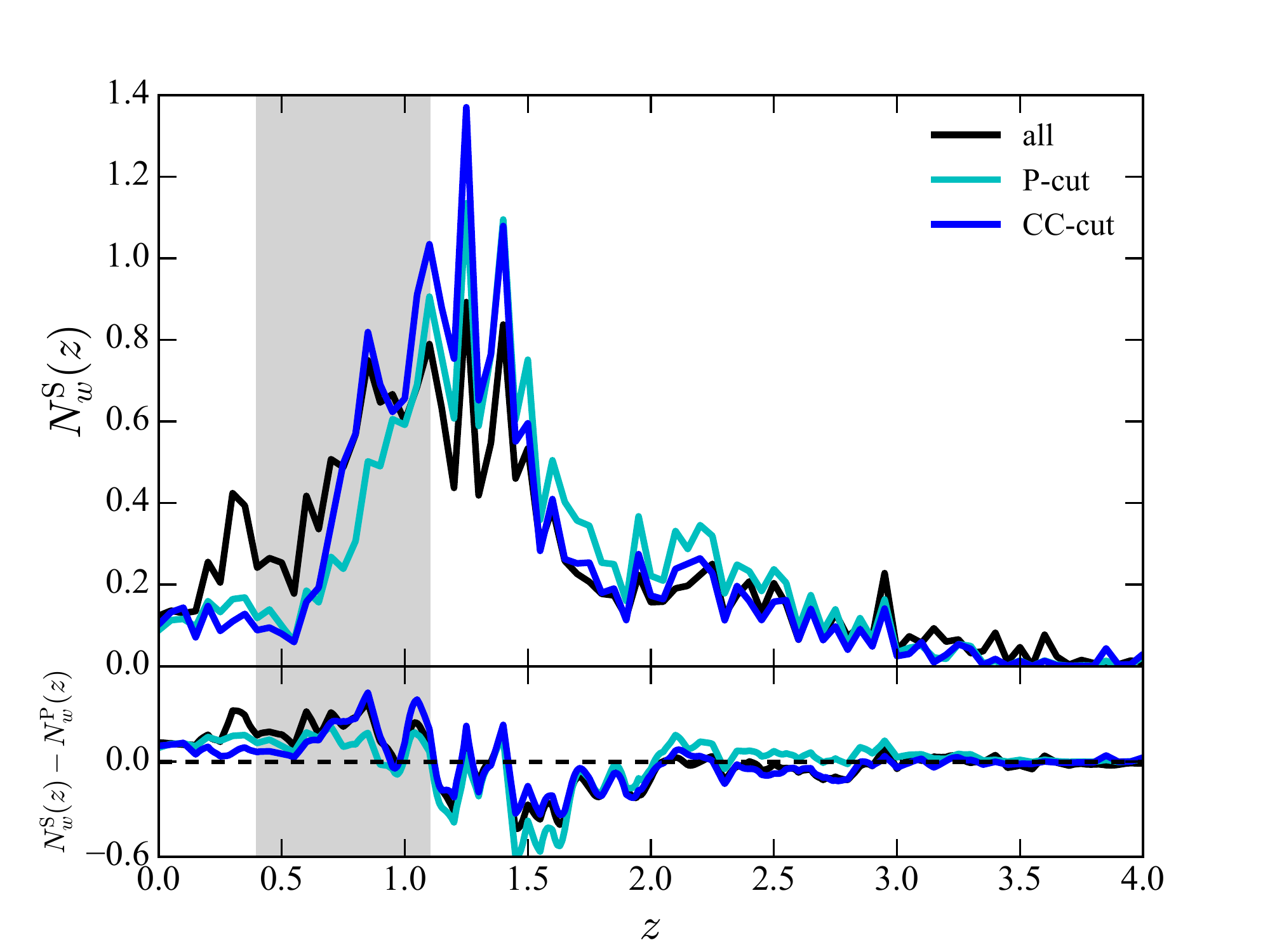}
\caption{Source redshift distributions, based on re-weighted spec-z's.  The bottom panels show the difference between the re-weighted spec-z distribution and the stacked $P(z)$ distribution (see text). Source samples are selected based on CC-cuts (blue) and P-cuts (cyan) and with no cuts (black). Lensing weights have been applied to each galaxy. The left panel shows the redshift distribution of sources behind low-z ($z_l<0.4$) clusters, and the right for high-z ($z_l\geq0.4$) clusters. For low-z clusters, the foreground contamination is negligible, $2-4\%$. For high-z clusters, the contamination is much higher, $12-22\%$ (see Table \ref{tab:contmlz}). }
\label{fig:wspecz_mlz}
\end{figure*}

We now attempt to estimate the reliability of the underlying redshift distribution from photo-z by comparing it with that derived from spec-z samples compiled in the HSC footprint \citep[for details of the spec-z surveys used see][]{Tanaka2017}. However,   spectroscopic samples are much shallower than those of photometric samples, and their color distribution may be very different since spectroscopic follow-up  typically misses  certain areas of color space \citep{Masters2015}. For these reasons, they may not be representative of the photometric sample and its redshift behavior. We can account for these differences by re-weighting the spec-z's according to their distribution in color and magnitude space \citep{Lima2008}. 
 
The \citeauthor{Lima2008} re-weighting method  assigns weights to galaxies in a spectroscopic subsample such that the weighted distributions of photometric observables,  e.g., multiband magnitudes, {colors and sizes,}
match those of the corresponding distributions of the photometric sample.
The weight is calculated as
\begin{equation}
  \label{eq:reweight}
  u_i =  \frac{\rho_i^{\rm p}(k)}{\rho_i^{\rm s}(k)},
\end{equation}
where $\rho_i^{\rm s}(k) \equiv k/V_i^{\rm s}(k)$ is a local density
of galaxies in  color, magnitude {and size} space. The density is defined
by the spherical volume centered on the $i$-th spec-z galaxy in which the
$k$ nearest neighbor galaxies are included. $\rho_i^{\rm p}$ is the
corresponding density defined in the same manner using the photometric
sample. We define the photometric sample as all the galaxies used for our analysis above, and located
within 3~\mpch~ from the CAMIRA cluster centers. We further calculate 
this for the P-cut and CC-cut constrained samples. We separate the clusters 
into two redshift bins, above and below $z_l=0.4$, as was done for our CC-selection 
analysis. This weight ensures that the
distribution of the spec-z sample in  magnitude and color space is
identical to that of the target photometric source sample. The effects of
incompleteness or large-scale structure are all absorbed in this
weight. The redshift distribution can then be estimated as, 
\begin{equation}
 N^{\rm S}(z_j)  
  =  \sum_{i}^{N_{\rm spec}} u_i(z_j) ,
\end{equation}
For each spec-z, we derive the weights of Equation~\ref{eq:reweight}. 
We further apply the lensing weight of each spec-z source, $w_{li}$, from Equation~\ref{eq:lensweight}  to the redshift distribution, 
\begin{equation}\label{eq:lensreweight}
 N^{\rm S}_w(z_j)
  =  \sum_{i}^{N_{\rm spec}}  {w}_{li} u_i(z_j) .
\end{equation}
Since the spec-z samples are not necessarily associated with
clusters, lens redshifts are randomly drawn from the redshift
distribution of the clusters. With this prescription, all the spec-z
samples can be used to evaluate the contamination rate due to foregrounds. In practice, 
we only a portion of the cross-validation spec-z
sample \citep[ID $= 5$; see][]{Tanaka2017}, since most of the cross-validation  samples (ID $=1-4$) are used to train and calibrate the photo-z and should therefore not be used to evaluate the
performance.

Figure~\ref{fig:wspecz_mlz} shows the  re-weighted spec-z
distributions, after applying the lensing weight as described above for  low-z (left) and high-z (right) clusters. 
For comparison, in the bottom panels of Figure~\ref{fig:wspecz_mlz} we also show the difference between the reweighted spec-z distribution and  the lensing-weighted stacked $P(z)$ distributions derived from {\sc mlz}. The weighted stacked $P(z)$ distribution is defined as
\begin{equation}\label{eq:stackedPz}
 N^{\rm P}_w(z_j)
   =  \sum_{i}^{N_{\rm p}} w_{li} P_i(z_j),
\end{equation}
where we use the lensing weight of Equation~\ref{eq:lensweight},
and the sum is taken over all  $N_{\rm p}$ galaxies in our photometric source sample  within 3$h^{-1}$Mpc
from the center of each cluster, applying corresponding cuts of either
CC- or P-cut. 

We define the fraction of foreground contamination in a source sample for a specific lens at redshift $z_{\rm l}$ as,
\begin{equation}
    f_{\rm FG}(z_{\rm l})    =    \frac{ \displaystyle{ \int_{0}^{z_{\rm l}} \!dz N^{\rm S}_w(z)}}%
    { \displaystyle{ \int_{0}^{\infty} \!dz N^{\rm S}_w(z)} }.
\end{equation}
For a lens sample with redshift distribution $p(z_{\rm l})$, the mean foreground fraction is computed as a weighted average of the redshift distribution,
\begin{equation}\label{eq:lenssum}
\langle f_{\rm FG} \rangle = \frac{\int {\rm d}z_{\rm l}\, p(z_{\rm l}) w(z_{\rm l}) f_{\rm FG} (z_{\rm l})}{\int {\rm d}z_{\rm l}\, p(z_{\rm l})  w(z_{\rm l}) },
\end{equation}
where the lens weight $w(z_{\rm l})$ is the sum of weights $w_{li}$ over all sources with respect to the lens redshift. We estimate errors from bootstrapping over the spec-z sample.
We  summarize the results  for each selection method in Table~\ref{tab:contmlz}, separately considering lenses above and below $z_l=0.4$. We find that  although the foreground contributions exist, they are small for the low-z case, typically $\lesssim4\%$ even without any cuts. For high-z lenses, the contamination is much higher, reaching $22\%$ without any cuts applied, but much improved for the P-cut and CC-cut methods, reaching only $\sim11\%$.

We furthermore estimate the photo-z calibration bias. This is defined as the ratio between the measured $\DSigma$ and the true $\DSigma_{\rm T}$ estimated from the reweighed spec-z's \citep[see][]{Mandelbaum2008b,Leauthaud2017},
\begin{equation}
b_z(z_l) +1 = \frac{\DSigma}{\DSigma_{\rm T}} = \frac{\sum_i w_{li} u_i\langle\Sigma_{\rm cr,P}^{-1}\rangle_i^{-1}/ \Sigma_{\rm cr,T,i}}{\sum_i w_{li} u_i}
\end{equation}
and $b_z$ gives the photo-z calibration bias. Here,  $\Sigma_{\rm cr,P},\ \Sigma_{\rm cr,T}$ are the photo-z and spec-z estimated critical densities, respectively. {When inserting the expression for the weight from Equation~\ref{eq:lensweight}, $w_{li} \propto \langle \Sigma_{\rm cr, P}^{-1}\rangle^2$, the sum in the numerator is a finite and well behaved quantity, proportional to  $\propto\Sigma_{\rm cr,T}^{-1}\langle\Sigma_{\rm cr,P}^{-1}\rangle$.}
Similar to the mean foreground fraction estimate (Equation~\ref{eq:lenssum}), we estimate the mean photo-z calibration bias by integrating over the redshift distribution of our cluster sample, for the low-z case $0.1<z_{\rm l}<0.4$, and the high-z case  $0.4\leq z_{\rm l}<1.1$. We present the results in Table~\ref{tab:fbias}. {For the low-z case, the bias is small, $4.5\%$  for the full sample, $1.9\%$  for the P-cut sample and only $1.4\%$  for the CC-cut sample. For  high-z clusters, the bias is significantly larger, estimated at $18\%$  for the full sample, while dropping to $11\%$  for the P-cut sample and $14\%$ for the CC-cut sample.} 
The overall agreement between the foreground fraction estimated above and the level of photo-z calibration bias indicates that indeed the foreground contamination  dominates the $\DSigma$ bias due to photo-z error (with only small contributions due to the scatter and bias in background galaxy photo-z). 

Overall, the P-cut and CC-cut selection methods succeed in mitigating  the bias  introduced by removing foregrounds, but the level is still high for high-z clusters. This bias should be corrected for when deriving masses of high-z clusters, after further improving the sampling of spec-z's in color and magnitude space.

There are a few important caveats regarding the validity of the re-weighting method worth mentioning: for one, this method is applicable only as far as the spec-z sample covers and samples all of the color/magnitude space of the source sample.  If some area of that space has no (or too few) spec-z galaxies, then this method  fails. When more complete spectroscopic campaigns are completed (e.g., \citealt{Masters2017} are targeting undersampled regions of color space to study the color-redshift relation) these estimates can be made more robustly. Another caveat is that we assume the spectroscopic completeness is independent of redshift, and even at fixed color and magnitude this may not be the case. 
This assumption is a known limitation of the re-weighting method, and has been so far adopted in other WL studies \citep[e.g.,][]{Hildebrandt2017}, albeit at much shallower survey depths. There are currently no studies in the literature that rigorously test the impact of this assumption for samples as deep as HSC. 
These assumptions may well contribute to a  systematic error that  exceeds the remaining uncertainty in the photo-z bias estimate, however, we are unable to reliably quantify it with the datasets currently at hand.

\begin{table}[t]
\tbl{Foreground Contamination Level, $f_{\rm FG}$ [\%] }{
{
\begin{tabular}{ccc}
\hline\hline
Method & $z<0.4$ & $z\ge0.4$ \\
\hline
All & $4.0\pm 0.3$ & $20\pm2 $ \\
P-cut & $2.3\pm0.3$ & $10\pm2$ \\
CC-cut & $2.8\pm0.4$ & $10\pm0.9$ \\
\hline
\end{tabular}
}
}
\label{tab:contmlz}
\end{table}

\begin{table}
\tbl{Photo-z Calibration Bias, $b_z$ [\%] }{
{
\begin{tabular}{ccc}
\hline
Method & $z<0.4$ & $z\ge0.4$ \\
\hline\hline
All & $-4.5\pm0.6$ & $-18\pm1$ \\
P-cut & $-1.9\pm0.5$ & $-11\pm1$ \\
CC-cut & $-1.4\pm0.5$ & $-14\pm2$ \\
\hline
\end{tabular}
}
}\label{tab:fbias}
\end{table}

\subsection{Boost Factors}
\label{subsec:boost}

With these careful selection methods  (P-cut and CC-cut), we have attempted to create restrictive source samples with minimal cluster contamination, so that no further correction for dilution is required. Figure~\ref{fig:gt_colorlim} and Figure~\ref{fig:gt_dzlim} should demonstrate that our selection is, by design, set to minimize the contamination -- we selected the cuts where the signal dilution is negligible. However, it is  useful to validate the residual contamination level independently. Boost factor corrections are typically used to account for any residual cluster contamination, by virtue of the radial correlation  of source galaxies compared to a field sample which is not expected to be related to the clusters. 
To accomplish this, one can either  stack the source sample around random points as the reference \citep{Sheldon2004}, or alternatively, stack ``fake" sources around the lens sample  (Murata et al., in prep, \citealt{Melchior2017}).  

For the former, a larger (thousands deg$^2$) survey area is ideal, in order for the correlation function to correctly yield a large-separation baseline \citep{Melchior2017}. We find that the boost factor derived this way displays  some non-typical behavior, where the number counts gradually declines toward the center starting at $\lesssim2~\mpch$, instead of rising  as expected for cluster contamination. To determine if this decline is caused by blending  with cluster galaxies, Murata et al. (2017, in prep) examine the blending effect using fake objects around CAMIRA clusters in HSC. They find this effect should only be significant below scales of $\sim 0.3~\mpch$, and cannot explain the behavior we see. We therefore  cannot utilize the boost factor derived this way at this stage.

For the latter, one would require a  fake source catalog, over the entire observed area, on which the same photometry, WL and photo-z cuts need to be applied, as well as the same color (in case of CC-cuts) or $P(z)$ cuts (in case of P-cut) for each selection method to be assessed. This is due to the non-trivial effect the color/photo-z selection may have on the number density of selected galaxies and which varies from field to field. Although \cite{Huang2017} and Murata et al. (2017, in prep) have implemented a fake object pipeline, \texttt{SynPipe}, it currently  does not carry the color and photo-z information needed for our test here.
We conclude an absolute boost factor cannot be derived in this way using the currently available products from \texttt{SynPipe}.
%

\begin{figure*}[tb]
\includegraphics[width=0.5\textwidth,clip]{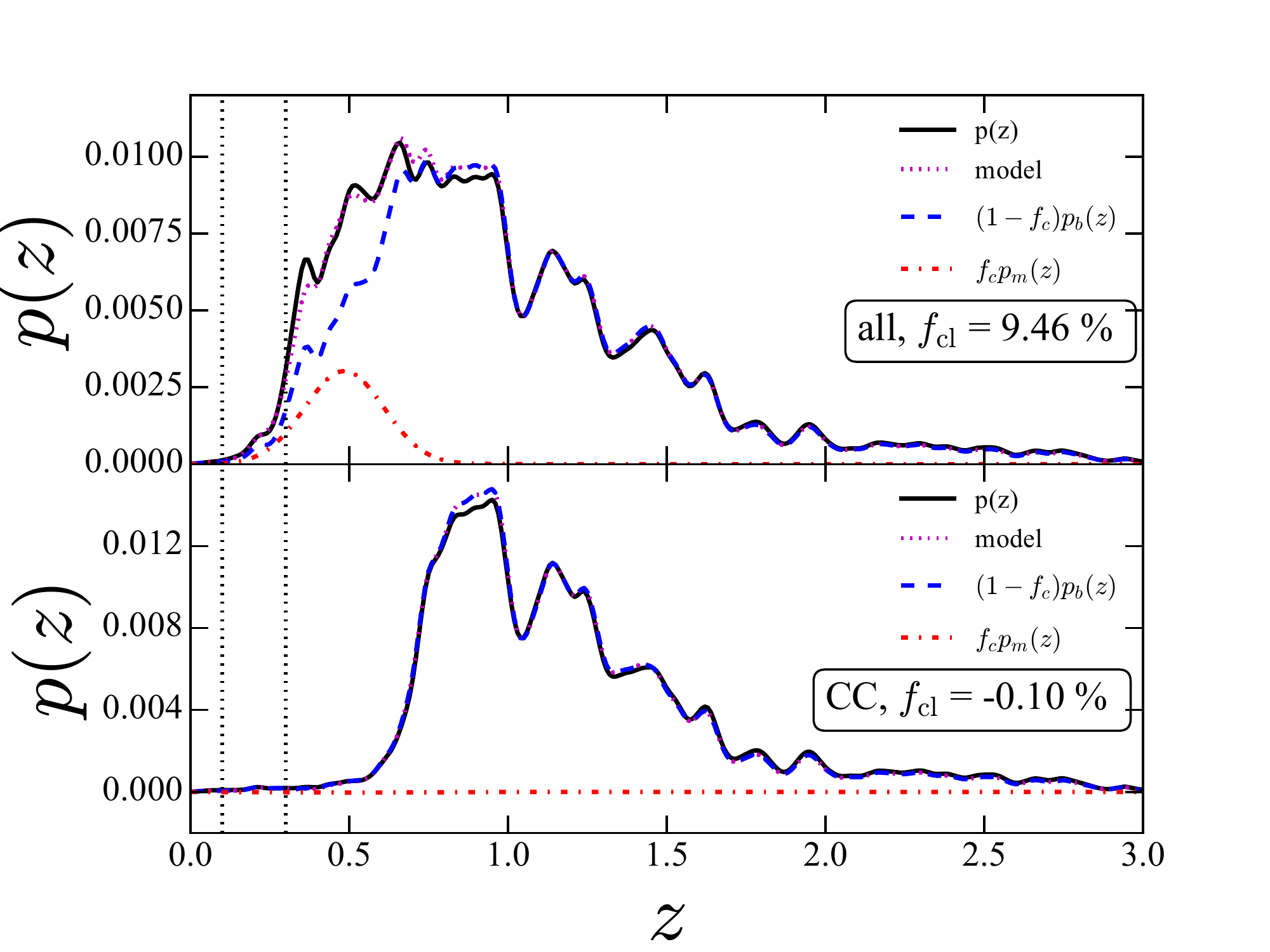}
\includegraphics[width=0.5\textwidth,clip]{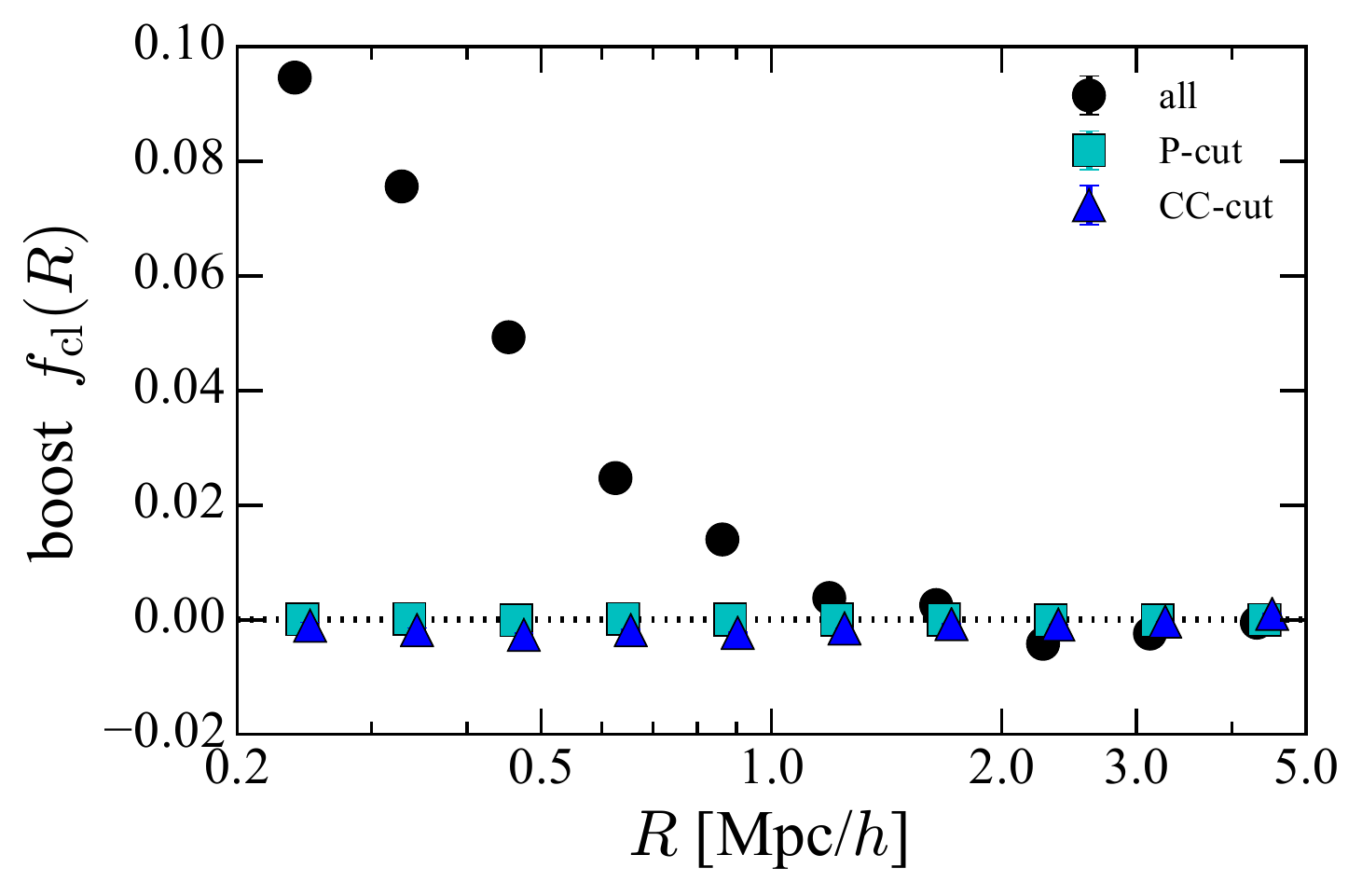}
\caption{Left: $p(z)$ decomposition in the first radial bin, $R\approx0.24$~\mpch, for sources behind clusters in the lens redshift range $0.1<z_{\rm cl}<0.3$. The black solid curves shows the mean $p(z)$ of the sources behind clusters and the blue dashed curve shows the mean $p(z)$ of sources around random points, i.e. the field $p_b(z)$. The red dashed-dotted line shows the  cluster member contribution, modeled as a Gaussian. The sum of the modeled cluster contribution and the field is given by the magenta dotted line and should overlap with the observed $p(z)$. The upper panel presents the decomposition of `all' sources (i.e., no cuts applied but the WL cuts), and the bottom panel  presents the decomposition of CC-cut sources, depicting no cluster contamination even at the inner radial bin.
Right: Boost factor profile, $f_{\rm cl}(R)$, based on the  $p(z)$ decomposition. Different curves show factors for different selection methods: all (black), P-cut (cyan) and CC-cuts (blue).  }
\label{fig:boost}
\end{figure*}

Instead, we attempt to estimate the boost factor by decomposing the redshift distribution of our source sample into a field galaxy component and a cluster member component. A variant of this method was first presented in \cite{Gruen2014}, and later adapted  for the DES cluster lensing analysis by \cite{Melchior2017}. For each method, in each radial bin, we measure the lensing-weighted mean photo-z $p(z)$ of sources around cluster positions,
\begin{equation}
p(z) = \frac{\sum_{l,s} w_{ls} p_s(z)}{\sum_{l,s} w_{ls}}.
\end{equation}
We furthermore measure the  lensing-weighted mean photo-z P(z) of sources around random positions, 
\begin{equation}
p_b(z) = \frac{\sum_{r,s} w_{rs} p_s(z)}{\sum_{r,s} w_{rs}}.
\end{equation}
which constitutes the "field" $p(z)$.
We then decompose the observed photo-z distribution $p(z)$ as a sum of the field distribution $p_b(z)$ and a cluster member contribution, $p_m(z)$,
\begin{equation}
p(z) = (1-f_{\rm cl})p_b(z)+f_{\rm cl}p_m(z).
\end{equation}
We model the cluster member distribution as a Gaussian distribution, jointly fitting its mean and width to all radial bins at once, while we fit the amplitude of the Gaussian, $f_{\rm cl}(R)$, in each radial bin independently. We show an example for this procedure in the upper left panel of Figure~\ref{fig:boost} using the full source sample (`all'), done for sources behind lenses in the redshift range $0.1<z<0.3$, for all richnesses, in the first radial bin, $R=0.24$~\mpch, where contamination is expected to be maximal. The contribution in this bin is estimated to be as high as $\sim9.5\%$. We repeat  this test for the P-cut and CC-cut samples, but fixing the Gaussian model mean and width values to those determined from the full sample and only fitting for the amplitude $f_{\rm cl}(R)$, since for those methods the cluster member contribution is expected to be negligible. This is  also evident in the bottom left panel of Figure~\ref{fig:boost}, showing the photo-z distributions in the first radial bin using CC-cut sources. There, the estimated cluster contamination is consistent with zero.
The resulting boost factor profile, $f_{\rm cl}(R)$, carried over all radial bins, is shown in the right panel of Figure~\ref{fig:boost}, for  `all' source (black), P-cut sources (cyan) and CC-cut sources (blue). 
As can be seen, the CC- and P-cut methods have curves consistent with zero down to scales of $0.3~\mpch$, suggesting there is no cluster contamination to within the 0.5\% uncertainties. The full sample (`all'), on the other hand, shows significant cluster contamination that is not corrected for by the photo-z information, starting at $\lesssim1~\mpch$~ and reaching as high as $9.5\pm0.1$ per cent contamination at the inner radial bin, $R\approx0.24$~\mpch. This demonstrates the strength of our applied selection methods in significantly removing  the cluster contamination from our source sample, and allowing us to derive the lensing signal on smaller scales.

We note, for sources behind $N_{\rm mem}>50$ clusters,  the mean $p(z)$ appears biased by the cluster contamination excess, as above, but also shows a dearth around $0.7<z<1$, compared to the field. We speculate this is caused by the intra-cluster light contaminating the photometry of $z\sim1$ blue galaxies, making them redder, and scattering them to lower redshifts. This may explain why  for high-richness clusters the P-cut method appears diluted compared to the CC-cut (in Figure 9).  It would then make this test inapplicable to high-richness clusters. However, our high-richness cluster sample is small, so we cannot explore this robustly. We leave this analysis to a future paper when the full HSC dataset is at hand.

\section{Discussion and Conclusions}
\label{sec:summary}

In this paper, we have investigated robust source selection techniques that separate the background galaxies used for cluster WL shape measurement and alleviate dilution of the lensing signal. 
We have compared different source selection schemes commonly used: (1) relying on photo-z's and their full PDFs to correct for dilution (all), (2) selecting background galaxies according to their colors (CC-cut), and (3) applying PDF cuts  so that  the cumulative PDF lies beyond the cluster redshift (P-cut). We explored the exact boundaries to set within these methods to derive the least contaminated background galaxy samples. We have used 912 CAMIRA clusters detected in HSC-SSP first-year data that span the redshift range $0.1<z<1.1$, allowing us to further explore these cuts for different  lens redshifts and  richnesses.

We demonstrate that simply relying on the photo-z PDFs  results in a cluster surface mass density profile that suffers from cluster dilution   at small radii. On the other hand, both the CC- and P-cut selection methods perform comparatively well in removing most of the cluster and foreground contamination, and are  consistent with each other.  Differences are only seen for the most rich clusters, $\ncor>50$, where the P-cut method produces a marginally diluted signal compared to the CC-cuts. However, our cluster sample is currently not large enough to demonstrate this dilution with high significance.  We have shown by virtue of modeling the signal with an NFW profile that not applying these cuts will result in a ($13\pm4$)\% underestimation of the mass, $M_{\rm 500c}$, and up to ($24\pm11$)\% underestimation of the concentration of the mass density profile,  $c_{\rm 500c}$.

We have attempted to verify these selection methods by applying them to reweighted spectroscopic samples, and find them to be consistent and largely free from foreground contamination to below the level of $2-4\%$ for low redshift lenses, $z_l<0.4$. At higher redshifts, $z_l\geq0.4$, the methods succeed in reducing the level of bias from  $20\%$ bias to $\lesssim10\%$. Although the source selection methods examined here greatly reduce the photo-z bias by removing foreground contaminants, the adopted reweighting methodology may still suffer from systematic uncertainties due to  spectroscopic  selection effects that may exceed the 1\% level.
By stacking the  source photo-z distributions for each method and comparing to the field, we have modeled the cluster contamination fraction to be quite significant for low-z clusters when no cuts are applied, reaching $9.5\%$ at $R\approx 0.24$~\mpch, yet being consistent with zero at all scales to within the 0.5\% uncertainties if applying the P-cut or CC-cut methods.

In summary, we conclude that applying either the P-cut or the CC-cut selection method is crucial for removing  cluster and foreground contamination and achieving an undiluted cluster mass profile. We note, however, that for very massive or nearby cluster samples, the conservative CC selection yields a more secure source sample with minimal contamination and less diluted  density profile.


\begin{ack}
We thank the anonymous referee for insightful comments that improved the manuscript. EM acknowledges fruitful discussions with Andy Goulding, Peter Melchior and Nick Battaglia.
The Hyper Suprime-Cam (HSC) collaboration includes the astronomical communities of Japan and Taiwan, and Princeton University. The HSC instrumentation and software were developed by the National Astronomical Observatory of Japan (NAOJ), the Kavli Institute for the Physics and Mathematics of the Universe (Kavli IPMU), the University of Tokyo, the High Energy Accelerator Research Organization (KEK), the Academia Sinica Institute for Astronomy and Astrophysics in Taiwan (ASIAA), and Princeton University. Funding was contributed by the FIRST program from Japanese Cabinet Office, the Ministry of Education, Culture, Sports, Science and Technology (MEXT), the Japan Society for the Promotion of Science (JSPS), Japan Science and Technology Agency (JST), the Toray Science Foundation, NAOJ, Kavli IPMU, KEK, ASIAA, and Princeton University. 
This paper makes use of software developed for the Large Synoptic Survey Telescope. We thank the LSST Project for making their code available as free software at  http://dm.lsst.org.
The Pan-STARRS1 Surveys (PS1) have been made possible through contributions of the Institute for Astronomy, the University of Hawaii, the Pan-STARRS Project Office, the Max-Planck Society and its participating institutes, the Max Planck Institute for Astronomy, Heidelberg and the Max Planck Institute for Extraterrestrial Physics, Garching, The Johns Hopkins University, Durham University, the University of Edinburgh, Queen?s University Belfast, the Harvard-Smithsonian Center for Astrophysics, the Las Cumbres Observatory Global Telescope Network Incorporated, the National Central University of Taiwan, the Space Telescope Science Institute, the National Aeronautics and Space Administration under Grant No. NNX08AR22G issued through the Planetary Science Division of the NASA Science Mission Directorate, the National Science Foundation under Grant No. AST-1238877, the University of Maryland, and Eotvos Lorand University (ELTE) and the Los Alamos National Laboratory.
Based [in part] on data collected at the Subaru Telescope and retrieved from the HSC data archive system, which is operated by Subaru Telescope and Astronomy Data Center at National Astronomical Observatory of Japan.
This paper makes use of packages available in Python's open scientific
ecosystem, including NumPy \citep{NumPy}, SciPy
\citep{SciPy},  matplotlib
\citep{matplotlib}, IPython \citep{IPython},
AstroPy \citep{astropy13}, and cluster-lensing \citep{Ford2016}.
The work reported on in this paper was substantially performed at the TIGRESS high performance computer center at Princeton University which is jointly supported by the Princeton Institute for Computational Science and Engineering and the Princeton University Office of Information Technology's Research Computing department.
HM is supported by the Jet Propulsion Laboratory, California Institute of Technology, under a contract with the National Aeronautics and Space Administration.
This work was supported in part by World Premier 
International Research Center Initiative (WPI Initiative), 
MEXT, Japan, and JSPS KAKENHI Grant Number 
26800093 and 15H05892.
KU acknowledges support from the Ministry of Science and Technology of Taiwan
through the grant MOST 103-2112-M-001-030-MY3.
RM is supported by the US Department of Energy Early Career Award Program.

\end{ack}


\appendix

\section{Color-Color Selection Recipe}
\label{app:CC}

Here we give a full description of the CC limits used in the CC selection method, along with the exact values chosen. We make use of the $g-i$ and $r-z$ colors,  using cModel magnitudes measured by the pipeline (see \citealt{Bosch2017} for the definition of cModel). In this CC space, we define a line which broadly follows the red-sequence of galaxies at $z\sim0.3$, seen as an overdensity in Figure~\ref{fig:CC}, 
\begin{equation}
{\rm CCseq}_\parallel= 2.276 \times  (r-z)  -0.152,
\end{equation}
and a perpendicular line as
\begin{equation}
{\rm CCseq}_\bot= 1/2.276 \times (r-z)  -0.152/2.276^2.
\end{equation}
We further define a line that follows the red-sequence in $r-z$ as a function of $z$ magnitude, as
\begin{equation}
rz{\rm seq}_\parallel = -0.0248 \times z +1.604.
\end{equation}

The red sample limits are then defined with respect to the above lines  as
\begin{eqnarray}
\label{eq:CCrlim1}
{\rm red}\Delta {\rm color}\#1& =& (g-i) - {\rm CCseq}_\parallel \\
\label{eq:CClim2}
{\rm red}\Delta {\rm color}\#2 & =& ((g-i) - {\rm CCseq}_\bot)/(1+1/2.276^2)
\end{eqnarray}
The specific cuts determined for each source sample (red or blue) at each lens redshift bin are explored in Section~\ref{subsec:CC} and are shown in Figure~\ref{fig:gt_colorlim} (dashed vertical lines).
For red sources associated with clusters at low redshifts, $z_l<0.4$, these definitions are
\begin{equation}
\begin{aligned}
{\rm red}\Delta {\rm color}\#1 & <  -0.7 &\  \&  \\
{\rm red}\Delta {\rm color}\#2 & <  4 &\  \&  \\
r-z & >  0.5  &\  \&  \\
z& > 21.
\end{aligned}
\end{equation}
For red sources associated with clusters at high redshifts, $z_l\geq0.4$, the cuts are 
\begin{equation}
\begin{aligned}
{\rm red}\Delta {\rm color}\#1 & <  -0.8 &\  \&  \\
{\rm red}\Delta {\rm color}\#2 & <  1.7 &\  \&  \\
r-z & >  0.5  &\  \&  \\
z& > 21.
\end{aligned}
\end{equation}

The blue source sample limits are  defined   as
\begin{equation}
{\rm blue}\Delta {\rm color}\#1 =(r-z) - rz{\rm seq}_\parallel 
\end{equation}
and the blue $\Delta {\rm color}\#2$ is the same as that defined for the red sample in Equation~\ref{eq:CClim2}.
The cuts used for blue sources associated with clusters at low redshifts, $z_l<0.4$, are 
\begin{equation}
\begin{aligned}
\big[\  {\rm blue}\Delta {\rm color}\#1 & <  -0.8  & |    &   &                     &        &\\
\big(\  {\rm blue}\Delta {\rm color}\#2 & <  0.5   & \&  & \ & g-i <  4 \big) &\ \big] & \&  \\
r-z & <  0.5 & & & & & \&  \\
z& > 22.
\end{aligned}
\end{equation}
The cuts used for blue sources associated with clusters at high redshifts, $z_l\geq0.4$, are 
\begin{equation}
\begin{aligned}
\big[\  {\rm blue}\Delta {\rm color}\#1 & <  -0.9  & |    &   &                     &        &\\
\big(\  {\rm blue}\Delta {\rm color}\#2 & <  0.3   & \&  & \ & g-i <  4 \big) &\ \big] & \&  \\
r-z & <  0.5 & & & & & \&  \\
z& > 22.
\end{aligned}
\end{equation}

\bibliographystyle{apj}
\bibliography{/Users/elinor/Dropbox/Documents/Elinor}

\end{document}